\documentclass[a4paper,11pt]{article}
\usepackage{jheppub} % for details on the use of the package, please see the JINST-author-manual
\usepackage{lineno}
\usepackage{ragged2e}
\usepackage{hyperref}
\usepackage{float}
\usepackage[T1]{fontenc} % if needed
\usepackage{amsfonts,amsmath,amssymb,bm,physics,slashed}
\usepackage{tikz}
\usetikzlibrary{arrows.meta,calc}
\usepackage{mathtools}
\usepackage{array} % validation-table and alphabet formatting
\usepackage{booktabs}

\usepackage[most]{tcolorbox}

\usepackage{subfigure}

\makeatletter
\gdef\@fpheader{}
\makeatother

\newcommand{\bR}{\mathbb R}

\newcommand{\bP}{\mathbb P}
\newcommand{\cA}{\mathcal A}

\renewcommand{\phi}{\varphi}

\newcommand{\Gr}{{\rm Gr}}

\usepackage{amsthm}

\theoremstyle{remark}

\theoremstyle{plain}

\theoremstyle{definition}

\title{\boldmath Deformed Amplituhedra and Interpolating Anomalous Dimensions on the Coulomb Branch}

\author[a]{Francesco Calisto,}
\author[b,c]{Subramanya Hegde,}
\author[d]{Mart\'in Lagares,}
\author[b,e]{Elia Mazzucchelli}

\affiliation[a]{Walter Burke Institute for Theoretical Physics and Leinweber Forum for Theoretical Physics,
California Institute of Technology, Pasadena, CA 91125, USA}

\affiliation[b]{Max Planck Institute for Physics,
Boltzmannstr.\ 8, 85748 Garching, Germany}

\affiliation[c]{School of Physics,
University of Hyderabad,
Hyderabad 500046, India}

\affiliation[d]{School of Mathematical Sciences,
Queen Mary University of London,
Mile End Road, London E1 4NS, United Kingdom}

\affiliation[e]{School of Natural Sciences,
Institute for Advanced Study,
Princeton, NJ 08540, USA}

\emailAdd{fcalisto@caltech.edu}
\emailAdd{subbu@uohyd.ac.in}
\emailAdd{m.lagares@qmul.ac.uk}
\emailAdd{eliam@ias.edu}

\abstract{
We study planar four-point scattering amplitudes on the Coulomb branch of \(\mathcal N=4\) super Yang--Mills theory, focusing on multi-parameter families of massive kinematics. We extend the deformed Amplituhedron construction of \cite{Arkani-Hamed:2023epq,Flieger:2025ekn} to general external kinematics characterized by a six-parameter embedding-space Gram matrix. We show that copositivity constraints on the inverse Gram matrix control the real geometry and restrict the numerator of the two-loop canonical form to the span of the two planar double-box structures expected for the scattering amplitude.
We then specialize to a three-parameter massive configuration and evaluate the relevant one- and two-loop integrals using differential equations directly in four dimensions, obtaining analytic expressions in terms of multiple polylogarithms. We investigate multiscale infrared limits of the proposed two-loop amplitude and find exponentiation through two loops. The kinematic family contains the limits governed by the cusp and octagon anomalous dimensions and interpolates between them through the walking anomalous dimension introduced in~\cite{Alday:2026cpk}. Our result reproduces the one-loop interpolation and fixes the two-loop coefficients in their proposed all-loop ansatz. Interestingly, our results suggest that the walking anomalous dimension may be entirely determined through all loop orders by the cusp and octagon anomalous dimensions.

}

\begin{document}

\maketitle

\allowdisplaybreaks
\section{Introduction}
\label{sec:introduction}

Scattering amplitudes of planar $\mathcal{N}=4$ super Yang--Mills (SYM) have been central to the development of modern amplitude methods over the past couple of decades \cite{Arkani-Hamed:2022rwr}. Amplitudes in this theory display remarkable properties such as dual conformal symmetry \cite{Drummond:2006rz,Drummond:2008vq} and Yangian symmetry \cite{Drummond:2009fd}, closely linked with their duality to null polygonal Wilson loops \cite{Alday:2007hr,Drummond:2007cf}. While loop amplitudes are plagued by infrared divergences, the divergent terms exponentiate and encode universal physical information, such as the cusp anomalous dimension \cite{Polyakov:1980ca,Korchemsky:1987wg,Bern:2005iz}. To study this IR exponentiation while maintaining dual conformal invariance, one approach is to introduce a mass regularization by computing IR-finite scattering amplitudes on the Coulomb branch of $\mathcal{N}=4$ SYM \cite{Alday:2009zm}. The exponentiation is then observed when taking the massless limit in the small-mass expansion \cite{Henn:2010bk,Henn:2010ir}.

More broadly, Coulomb-branch observables have recently attracted interest in connection with integrability, Regge physics, and strong-coupling and string-theoretic limits. There has been recent progress on the integrability of the Coulomb-branch theory beyond scattering amplitudes \cite{Coronado:2025xwk,Demjaha:2025axy}. Using dual conformal invariance, one can relate the cusp anomalous dimension to Regge trajectories and extract the bound-state spectrum \cite{Caron-Huot:2014gia}. It has further been observed that Coulomb-branch amplitudes interpolate to open-string amplitudes at strong coupling, and computing loop amplitudes for new Coulomb-branch configurations can provide data for consistent string amplitudes beyond the well-known Virasoro--Shapiro and Veneziano amplitudes \cite{Alday:2025pmg}. Thus, there is growing interest in computing loop amplitudes for novel Coulomb-branch configurations and in understanding the physical information encoded by their infrared behavior.

The Coulomb branch of $\mathcal{N}=4$ SYM is reached when scalar expectation values move away from the origin of the moduli space and thereby spontaneously break the gauge group and R-symmetry. The simplest Coulomb-branch configuration studied in \cite{Alday:2009zm} involves the gauge-group breaking
$U(M+N)\rightarrow U(1)^M\times U(N)$, with $M\ll N$, and R-symmetry breaking $SO(6)\rightarrow SO(5)$, and can be regarded as amplitudes of ten-dimensional $\mathcal{N}=1$ SYM with momenta restricted to five spacetime dimensions. While this setup captures the IR exponentiation associated with the cusp anomalous dimension, other Coulomb-branch configurations can encode information about universal physical quantities such as the octagon anomalous dimension \cite{Coronado:2018cxj,Caron-Huot:2021usw}. At finite coupling, these anomalous dimensions are known from integrability-based approaches \cite{Beisert:2006ez,Belitsky:2019fan}. An interpolation between the cusp and octagon anomalous dimensions is also known from integrability \cite{Basso:2020xts}. More recently, in~\cite{Alday:2026cpk} the authors proposed a Coulomb-branch configuration in which loop amplitudes exhibit an infrared divergence governed by a \emph{walking anomalous dimension}, providing a different interpolation between the cusp and octagon anomalous dimensions. It is this mass configuration and the corresponding anomalous dimensions that form one of our main objects of study in this paper.

A natural question is whether these more general Coulomb-branch amplitudes admit a geometric formulation comparable to the Amplituhedron description at the conformal point. At the origin of the moduli space of planar $\mathcal{N}=4$ SYM, the Amplituhedron provides a geometric formulation that encodes the Yangian symmetry of tree-level amplitudes and loop integrands \cite{Arkani-Hamed:2013jha,Arkani-Hamed:2010zjl,Arkani-Hamed:2012zlh,Ferro:2016zmx}. Away from the origin of the moduli space, besides dual conformal invariance, Coulomb-branch amplitudes have been known to exhibit recursive properties at tree level \cite{Elvang:2011ub,Herderschee:2019dmc}, as well as structured loop-level leading singularities \cite{MdAbhishek:2023nvg}. Recent works have also established a connection between tree-level Coulomb-branch amplitudes and the symplectic Grassmannian \cite{Cachazo:2018hqa,Bering:2022tdr,Cortes:2025xmu}. A complementary approach is to ask how the Amplituhedron geometry itself must be modified in order to describe amplitudes on the Coulomb branch. Such a proposal was introduced in~\cite{Arkani-Hamed:2023epq,Flieger:2025ekn}, where the inequalities defining the Amplituhedron were deformed at one and two loops to obtain loop integrands for a particular Coulomb-branch vacuum configuration. This raises the more immediate question of how far the deformed-Amplituhedron construction extends when the Coulomb-branch mass configuration is generalized\footnote{For the associahedron description of planar scalar theories, such a program was initiated earlier in~\cite{Jagadale:2022rbl}.}.

The geometry of the deformed Amplituhedron is also interesting in its own right from a mathematical perspective. Positive geometries provide a framework in which distinguished differential forms are determined by the boundary structure of an underlying real geometry~\cite{Arkani-Hamed:pos_geom,Lam:2022yly,Ranestad:2025positive}, with the Amplituhedron furnishing one of their central examples~\cite{Arkani-Hamed:2013jha,Arkani-Hamed:2013kca}. The deformed Amplituhedron introduced in~\cite{Arkani-Hamed:2023epq} gives a particularly natural variation of this construction. In the Grassmannian description, the special Schubert divisors defining the undeformed one-loop geometry are replaced by more general hyperplane sections of the Grassmannian~\cite{Mazzucchelli:2025hyperplane}. Already at one loop this picture has a very simple realization: the massive four-point box is encoded by a positive geometry in $\mathrm{Gr}(2,4)$ bounded by such deformed hyperplane sections. The relevant problem is therefore not only algebraic but also real-geometric: one must determine which such arrangements bound a consistent positive region. In our construction this question is naturally related to copositivity properties of the embedding-space Gram matrix.

Among the main goals of this paper is understanding how this geometry extends to two loops for more general mass configurations compatible with the embedding-space formalism. At two loops, two one-loop geometries are coupled by a mutual-positivity condition, and the resulting boundary structure constrains the allowed logarithmic form. This may be viewed as the deformed counterpart of the geometric analysis of the ordinary two-loop Amplituhedron carried out in~\cite{Dian:2024hil}. The geometric construction naturally allows six independent normalized Gram entries, while the explicit amplitude calculation considered below is restricted to a three-parameter Coulomb-branch subfamily with a direct mass interpretation. This family is sufficiently general to capture the interpolation of interest while remaining simple enough for the loop integrals to be treated analytically.

We summarize here our main contributions.

\paragraph{Deformed geometries and their integrands.}
We extend the deformed-Amplituhedron construction of~\cite{Arkani-Hamed:2023epq,Flieger:2025ekn} beyond its original two-parameter mass configuration. The external massive kinematics are encoded by an embedding-space Gram matrix with six independent entries. We show that copositivity properties of its inverse provide a criterion for determining which residual loci can intersect the real positive region and therefore constrain the boundary structure of the deformed Amplituhedron. At one loop, the resulting positive geometry has the box integrand as its canonical logarithmic form. At two loops, we derive explicit copositivity conditions under which the allowed logarithmic form is restricted to the span of the two planar double-box integrands, reproducing the expected integral topologies of the Coulomb-branch amplitude. We expect these copositivity conditions to play an important role in extending the construction to higher loops.

\paragraph{Analytic amplitudes and infrared behavior.}
We compute the one-loop amplitude and the two-loop double-box integrals explicitly in the three-parameter family introduced above. Assuming the double-box decomposition motivated by the deformed geometry and by its known \(z=0\) limit, these results determine the corresponding proposed two-loop amplitude. This kinematic family admits a physical realization with all external particles of the same mass $M$ and all massive internal propagators of a common mass $m$, while the propagators joining distinct loop lines remain massless. The corresponding box and double-box integrals are evaluated directly in four dimensions using finite differential equations (DEs), and we obtain analytic expressions in terms of Goncharov Polylogarithms (GPLs).

We use these results to study the infrared behavior of the amplitude within the three-parameter family. We consider the walking high-energy limit of \cite{Alday:2026cpk}, in which the hard invariants and the two mass scales $M$ and $m$ scale at different rates.
For the proposed two-loop amplitude, its logarithm exhibits the expected cancellation of quartic and cubic logarithmic divergences, leaving a leading double-logarithmic behavior determined by the walking anomalous dimension introduced in \cite{Alday:2026cpk}. This exponentiation provides an important consistency check of the assumed two-loop double-box normalization.
We compute the walking anomalous dimension up to two loops, reproducing the one-loop interpolation of \cite{Alday:2026cpk} and determining the two-loop coefficients in their all-loop amplitude ansatz. Our result shows that the undetermined constants $\gamma$ and $\tilde \gamma$ coincide through two loops and correspond to half the cusp anomalous dimension, suggesting that the all-loop walking anomalous dimension is completely fixed by the cusp and octagon anomalous dimensions.

\bigskip

The paper is organized as follows. Section~\ref{sec: Coulomb branch review} reviews the Coulomb-branch configurations relevant for our construction and introduces the three-parameter family studied explicitly in the paper. Section~\ref{sec:deformed-Amplituhedron} develops their geometric description in terms of deformations of the Amplituhedron and analyzes the boundary and residual structure that constrains the one- and two-loop integrands. In Section~\ref{sec:three-parameter-amplitudes}, we evaluate the resulting box and double-box integrals directly in four dimensions and obtain analytic expressions in terms of multiple polylogarithms. Section~\ref{sec:walking} studies their high-energy limits and extracts the corresponding walking anomalous dimension, including its interpolation between the cusp and octagon regimes and the comparison with the form-factor result. Finally, Section~\ref{sec:conclusions} summarizes our results and discusses several directions for future work. We also include several appendices and ancillary files that complement the results presented
in the main body of the paper.

\section{\texorpdfstring{The Coulomb branch of ${\cal N}=4$ super Yang--Mills}{The Coulomb branch of N=4 super Yang--Mills}}
\label{sec: Coulomb branch review}

As described in the Introduction, in this paper we will study the geometric description and IR properties of a rich family of massive Coulomb-branch amplitudes. With this goal in mind, we find it instructive to begin by presenting in this section a brief review of the main characteristics of the Coulomb branch of the ${\cal N}=4$ super Yang--Mills (SYM) theory. At the origin of its moduli space, the action of the ${\cal N}=4$ SYM theory with gauge group $U(N+4)$ is written as~\cite{Brink:1976bc}
\begin{equation}
\int d^4x \, {\rm Tr} \left(-\frac{1}{4} {\bf F}_{\mu \nu}^2 -\frac{1}{2} (D_{\mu} {\bf \Phi}^I)^2 + \frac{g_{\rm YM}^2}{4} [{\bf \Phi}^I, {\bf \Phi}^J]^2 + \frac{i}{2} \bar{{\bf \Psi}} \Gamma^{\mu} D_{\mu} {\bf \Psi} \right. \nonumber  \; \; \left.+ \frac{g_{\rm YM}}{2} \bar{{\bf \Psi}} \Gamma^I [{\bf \Phi}_I,{\bf \Psi}] \right) \,,
\label{N=4 sYM action-origin}
\end{equation}
Here $D_{\mu}=\partial_{\mu}-i g_{\rm YM}[{\bf A}_{\mu},\cdot]$, with $\mu=0,\dots,3$ and $I=1,\ldots,6$. The fields ${\bf A}_{\mu}$ are the $U(N+4)$ gauge fields, ${\bf \Psi}$ is a ten-dimensional Majorana--Weyl fermion, ${\bf \Phi}^I$ are six real scalar fields transforming in the vector representation of the $SO(6)$ R-symmetry group and the set $\{ \Gamma^{\mu}, \Gamma^I \}$ spans the ten-dimensional Clifford algebra; the six internal matrices are labelled by $I=1,\ldots,6$, independently of the spacetime index $\mu$. All the fields transform in the adjoint representation of the $U(N+4)$ gauge group. The gauge coupling is denoted by $g_{\rm YM}$, and in what follows we will use the notation $g^2=\frac{g^2_{YM}N}{16 \pi^2}$ for the 't Hooft coupling.

The Coulomb branch of ${\cal N}=4$ SYM is reached by expanding the action \eqref{N=4 sYM action-origin} around saddle-point configurations in which the ${\bf \Phi}^I$ scalar fields take constant non-zero values. In particular, we will be interested in cases in which the gauge symmetry is spontaneously broken to $U(N+4)\to U(N) \times U(1)^4$. These configurations are reached by considering saddle-point solutions to the equations of motion in which the scalar fields behave as
\begin{align}
    {\bf \Phi}^I_{0} &= \left( \begin{array}{cc}
       0_{N\times N}  & 0_{N\times 4}  \\
    0_{4\times N} & \frac{1}{g_{\rm YM}} \, {\rm diag}(y^I_1, y^I_2, y^I_3, y^I_4)  
    \end{array}\right) \,, 
\label{symmetry breaking pattern}
\end{align}
for some $SO(6)$ vectors $y_i^I$, $i=1, \dots, 4$, while the gauge and spinor fields vanish.
The vectors $y_i$ denote the positions relative to the coincident $SU(N)$ block. 

Let us now turn to the analysis of the mass terms that appear in the Coulomb-branch action. To that end it proves convenient to write the adjoint fields as
\begin{align}
 {\bf \Phi}^I &=  \left( \begin{array}{cc}
       \phi^I_{\mathsf{a}\mathsf{b}}  & \alpha^I_{\mathsf{a}i}  \\
    \alpha^{\dagger I}_{i \mathsf{b}} & \kappa^I_{ij} 
    \end{array}\right) \,, \qquad 
    &{\bf \Psi}&= \left( \begin{array}{cc}
       \psi_{\mathsf{a}\mathsf{b}}  & \zeta_{\mathsf{a}i}  \\
    \eta_{i \mathsf{b}} & \theta_{ij} 
    \end{array}\right) \,, 
    &{\bf A}_{\mu}&= \left( \begin{array}{cc}
       A_{\mu,\mathsf{a}\mathsf{b}}  & W_{\mu,\mathsf{a}i}  \\
    W^{\dagger}_{\mu,i\mathsf{b}} & V_{\mu, ij} 
    \end{array}\right) \,.  
\label{CB field decomposition}
\end{align}
with $i,j=1,\dots,4$ and $\mathsf{a},\mathsf{b}=1, \dots, N$. Then, after adding an appropriate gauge-fixing term (see \cite{Alday:2009zm}), the mass terms of the Coulomb-branch action~\eqref{N=4 sYM action-origin} become
\begin{equation}
\begin{aligned}
\int d^4x \, \Bigg[&
- \sum_{\mathsf{a}=1}^N \sum_{i=1}^4
y_i^2
\left(
(W_\mu^\dagger)_{i\mathsf{a}} W^\mu_{\mathsf{a}i}
+(\alpha^{I\dagger})_{i\mathsf{a}}\alpha^I_{\mathsf{a}i}
\right)
- \sum_{i,j=1}^4
\frac{(y_i-y_j)^2}{2}
\left(
(V_\mu)_{ij}V^\mu_{ji}
+\kappa^I_{ij}\kappa^I_{ji}
\right)
\\
&+ \frac{1}{2}\sum_{\mathsf{a}=1}^N \sum_{i=1}^4
y_i^I
\left(
\bar{\eta}_{\mathsf{a}i}\Gamma^I\eta_{i\mathsf{a}}
-\bar{\zeta}_{i\mathsf{a}}\Gamma^I\zeta_{\mathsf{a}i}
\right)
- \frac{1}{2}\sum_{i,j=1}^4
(y_i-y_j)^I\,
\bar{\theta}_{ij}\Gamma^I\theta_{ji}
\Bigg] \, .
\end{aligned}
\label{mass terms CB}
\end{equation}
We therefore see that the fields $A_{\mu}, \varphi^I$ and $\psi$ are massless, while the fields $W_{\mu,\mathsf{a}i}$ and $\alpha^I_{\mathsf{a}i}$ have mass $m_{i}^2=y_{i}^2$ and the fields $V_{\mu,ij}$ and $\kappa^I_{ij}$ have mass $M_{ij}^2=(y_{i}-y_{j})^2$.

As described in~\cite{Alday:2009zm}, there is a well-understood holographic dual of the Coulomb branch of ${\cal N}=4$ SYM.
Starting from the origin of the moduli space of ${\cal N}=4$ SYM with gauge group $U(N+4)$, whose holographic dual can be famously obtained by considering the near-horizon limit of a stack of coincident $N+4$ D3-branes in type IIB string theory~\cite{Maldacena:1997re}, one can holographically describe Coulomb-branch configurations such as \eqref{symmetry breaking pattern} by placing four of the aforementioned D3-branes at a finite distance from the others. 
In the near-horizon limit, this can be interpreted as placing four D3-branes in the $AdS_5 \times S^5$ background where the dual string theory lives. The positions of the branes along the holographic direction precisely determine the masses of the off-diagonal excitations $W_{\mu},\,  \alpha^I, \eta$ and $\zeta$. Similarly, the separation between the branes (both in $AdS_5$ and in $S^5$) determines the masses of the diagonal excitations $V_{\mu},  \, \kappa^I$ and $\theta$.

A key property of Coulomb-branch amplitudes, which will be of relevance in our analysis, is their $SO(2,4)$ dual conformal invariance (DCI). This symmetry can be understood by considering the holographic construction of the amplitudes, as initially pointed out in~\cite{Alday:2009zm}. Crucially, the $AdS_5 \times S^5$ background is self-dual under T-duality, turning the D3-branes that give rise to massive Coulomb-branch excitations into D-instantons. In this T-dual picture, the massive Coulomb-branch amplitudes can be computed in their strong-coupling limit as the minimal area of a string that ends at the four D-instantons. The dual conformal invariance of the amplitude is precisely the $SO(2,4)$ invariance of this string under the isometries of the T-dual $AdS_5$ spacetime. We should stress that the D-instantons are located at finite positions along the holographic direction, rendering a finite scattering amplitude. In the massless limit the D-instantons move towards the boundary of $AdS_5$, describing the vertices of a light-like polygon and giving rise to the well-known polygonal Wilson loop/scattering amplitude duality~\cite{Alday:2007hr,Alday:2007he,Drummond:2007aua,Brandhuber:2007yx,Drummond:2007cf}.

\subsection{One- and two-loop amplitudes}\label{sec:onetwoloopamp}

In this paper we will focus on the scattering amplitude
\begin{equation}
{\cal A} \left[\kappa^I_{12} (p_1) \kappa^J_{23} (p_2) \kappa^I_{34} (p_3) \kappa^J_{41} (p_4) \right] \,, \quad  I\neq J \,, \quad I,J \in \{ 5, 6\} \,,
\label{kappa amplitude}
\end{equation}
between four massive scalar fields in the $U(1)^4$ sector, where no summation over the $I$ and $J$ indices is implied. For simplicity we will further assume that the $y_{i}^I$ vectors live in the hyperplane spanned by the $I=1, \dots, 4$ directions, i.e. 
\begin{align}
y_{i}^{I}=0\,, \quad {\rm for} \quad  I \in \{5,6\}\,.
\end{align}

We will work in the 't Hooft limit of the theory, and we will write the perturbative expansion in the 't Hooft coupling $g$ as
\begin{equation}\label{eq:cbampdef}
\mathcal M \equiv \frac{\mathcal A}{\mathcal A_{\mathrm{tree}}}
=1+g^2\mathcal M^{(1)}+g^4\mathcal M^{(2)}+\mathcal O(g^6) \,,
\qquad g^2=\frac{g_{\mathrm{YM}}^2N}{16\pi^2},
\end{equation}
where we are normalizing the amplitude ${\cal A}$ by its tree-level value ${\mathcal A_{\mathrm{tree}}}$. In this limit a major simplification occurs. To see this, one should recall that in the Coulomb-branch setup defined by the vacuum expectation value \eqref{symmetry breaking pattern} one can distinguish between two types of $U(N+4)$ color indices. On the one hand, one has $\mathsf a,\mathsf b,\dots=1,\dots,N$ indices associated with the unbroken $U(N)$ block. On the other hand, one has $i,j,\dots=1,\dots,4$ indices associated with the four separated $U(1)$ factors. When using the double-line notation introduced by 't Hooft to write gauge-theory Feynman diagrams, one can have propagators with two unbroken gauge lines (for the $A_{\mu,\mathsf{a}\mathsf{b}}, \, \psi_{\mathsf{a}\mathsf{b}}$ and $\varphi^I_{\mathsf{a}\mathsf{b}}$ fields), two broken gauge lines (for the $V_{\mu,ij}, \, \theta_{ij}$ and $\kappa^I_{ij}$ fields), or one broken and one unbroken line (for the off-diagonal fields $W_{\mu,\mathsf{a}i}, \, \zeta_{\mathsf{a}i}, \, \eta_{i\mathsf{a}}$ and $\alpha^I_{\mathsf{a}i}$). It is easy to see then that in the planar limit the diagrammatic expansion of observables is dominated by diagrams with loops of unbroken gauge indices. 
For the case of the amplitude \eqref{kappa amplitude}, leading-order graphs in the $1/N$ expansion will be characterized by an external line of broken gauge indices, with all internal lines corresponding to the unbroken gauge group. 

As presented in~\cite{Alday:2009zm}, a standard perturbative computation shows that at one loop 
the sum over the contributing Feynman diagrams 
receives contributions only from box topologies. Omitting the dependence on the masses, this results in the equation 
\begin{align}
    \mathcal M^{(1)}(s,t) &= - \mathcal{B}^{(1)}(s,t) \, , \label{eq:1loopampdef}
\end{align}
where we defined the one-loop box integral as
\begin{align}
\mathcal{B}^{(1)}(s,t)
&=\int\frac{\mathrm d^4k_1}{i\pi^2}\,
\frac{( ( y_{1}-y_3)^2-s ) ( ( y_{2}-y_4)^2-t )}{
[k_1^2+y_1^2]
[(k_1+p_1)^2+y_2^2]
[(k_1+p_1+p_2)^2+y_3^2]
[(k_1-p_4)^2+y_4^2]} \,,\label{eq:1loopbox}
\end{align}
with
\begin{align}
p_{i}^2 = -(y_{i}-y_{i+1})^2 \,, \qquad 
s = -(p_1+p_2)^2 \,, \qquad
t = -(p_2+p_3)^2 \,.
\end{align}
Similarly, at two loops it is expected that the amplitude has a double-box decomposition
\begin{align}
    \mathcal M^{(2)}(s,t) &= \mathcal{B}_s^{(2)}(s,t) + \mathcal{B}_t^{(2)}(s,t)\,, \label{eq:2loopampdef}
\end{align}
where the double-box $s$-channel integral is
\begin{align}
\mathcal{B}^{(2)}_s(s,t)
&=\int\frac{\mathrm d^4k_1\,\mathrm d^4k_2}{(i\pi^2)^2}\,
\frac{((y_1-y_3)^2-s)^2 ((y_2-y_4)^2-t)}{
[k_1^2+y_1^2]
[(k_1+p_1)^2+y_2^2]
[(k_1+p_1+p_2)^2+y_3^2]}
\nonumber\\[-1pt]
&\hspace{8mm}\times
\frac{1}{
[(k_1-k_2)^2]
[k_2^2+y_1^2]
[(k_2+p_1+p_2)^2+y_3^2]
[(k_2-p_4)^2+y_4^2]} \,,
\label{B2}
\end{align}
and $\mathcal{B}^{(2)}_t(s,t)=\mathcal{B}^{(2)}_s(t,s)$. Such a two-loop double-box structure has been found for certain Coulomb-branch configurations in \cite{Abhishek:2023lva,Arkani-Hamed:2023epq,Flieger:2025ekn}, and we conjecture here its extension to a larger family of configurations. The kinematic normalizations in the numerator of~\eqref{eq:1loopbox} and~\eqref{B2} ensure that the integrals are dual conformal invariant, as explained below.

Let us note that when using four-dimensional dual-space coordinates
\begin{align}
p_i= x_{i+1}-x_i \, , \qquad k_1=x_{1}-x_{5} \,, \qquad k_2=x_{1}-x_{6} \,, 
\end{align}
and the corresponding ten-dimensional coordinates
\begin{align}
\hat{x}_i& := \left(x_i, y_{i} \right)\,, \qquad 
\hat{x}_5 := (x_5,0) \,, \qquad 
\hat{x}_6 := (x_6,0) \,,
\end{align}
the one-loop amplitude \eqref{eq:1loopampdef} and the proposed two-loop amplitude \eqref{eq:2loopampdef} become
\begin{align}
{\cal M}^{(1)}&= 
-\int 
\frac{d^4x_5}{i \pi^2} \, \frac{\hat{x}_{13}^2 \, \hat{x}_{24}^2}{\hat{x}_{15}^2\hat{x}_{25}^2\hat{x}_{35}^2\hat{x}_{45}^2} \,,
\label{one loop amplitude dual space} \\
{\cal M}^{(2)}&= 
\int 
\frac{d^4x_5}{i \pi^2} \, \frac{d^4x_6}{i \pi^2} \, \left( \frac{\hat{x}_{13}^4 \, \hat{x}_{24}^2}{\hat{x}_{15}^2\hat{x}_{25}^2\hat{x}_{35}^2 \hat{x}_{56}^2 \hat{x}_{16}^2\hat{x}_{36}^2\hat{x}_{46}^2} + \frac{\hat{x}_{13}^2 \, \hat{x}_{24}^4}{\hat{x}_{15}^2\hat{x}_{25}^2\hat{x}_{45}^2 \hat{x}_{56}^2 \hat{x}_{26}^2\hat{x}_{36}^2\hat{x}_{46}^2}  \right)\,,
\label{two loop amplitude dual space}
\end{align}
with $\hat{x}_{ij}:=\hat{x}_i-\hat{x}_j$, where we use the ten-dimensional Minkowski norm.

\subsection{Coulomb-branch configurations of interest}
\label{sec: CB configs of interest}

From now on we will work with Coulomb-branch amplitudes of the form \eqref{kappa amplitude} and we will impose that
the $SO(6)$ inner products between non-adjacent \(y_i\) vectors vanish, namely\footnote{This condition will be important in Section \ref{sec:deformed-Amplituhedron} when showing that the integrands \eqref{one loop amplitude dual space} and \eqref{two loop amplitude dual space} can be rewritten using the six-dimensional embedding-space formalism to be introduced in Section \ref{sec:embedding-formalism}.} 
\begin{equation}
y_1\!\cdot y_3 = 0 \, , \qquad
y_2\!\cdot y_4 = 0 \, .
\label{y1y3 y2y4 condition} 
\end{equation}
The scattering amplitude is then generically described by internal and external masses,
\begin{equation}
m_i^2 :=y_i^2 \,, \qquad M_i^2 :=-p_i^2=(y_i-y_{i+1})^2=m_i^2+m_{i+1}^2-2y_i \cdot y_{i+1}\,.
\end{equation}
As we will discuss in detail in Section \ref{sec:embedding-formalism}, under the condition \eqref{y1y3 y2y4 condition} the integrands \eqref{one loop amplitude dual space} and \eqref{two loop amplitude dual space} are invariant under $SO(2,4)$ dual conformal transformations and $GL(1)^4$ projective rescalings.
This allows one to express a generic amplitude of the form \eqref{kappa amplitude} under the condition \eqref{y1y3 y2y4 condition} in terms of six cross ratios invariant under dual conformal transformations and projective rescalings, 
which as we will show can be chosen to be 
\begin{align}
 {\cal G}_{13}&= \frac{-s+m_1^2+m_{3}^2}{2m_1 m_{3}} \,, &{\cal G}_{24}&= \frac{-t+m_2^2+m_{4}^2}{2m_2 m_{4}} \,, &{\cal G}_{i,i+1}&=\frac{-M^2_i+m_i^2+m_{i+1}^2}{2m_i m_{i+1}}=\frac{y_i \cdot y_{i+1}}{m_i m_{i+1}} \,. \nonumber
\end{align}
In Section \ref{sec:embedding-formalism} we show that the above cross-ratios precisely correspond to the entries of the normalized Gram matrix ${\cal G}$ of the four dual momentum vectors in embedding space, which are completely determined by the Mandelstam variables $s$ and $t$ and the choice of $SO(6)$ vectors $y_i$ (via their norms $y_i^2$ and their adjacent products $y_i \cdot y_{i+1}$).

In Sections \ref{sec:three-parameter-amplitudes} and \ref{sec:walking} we focus on the particular case in which for every $i=1,\dots,4$ 
\begin{align}
    M^2_i=M^2 \,, \qquad m^2_i=m^2  \,.
\label{three-parameter condition}
\end{align}
Coulomb-branch setups of this kind are described by only three independent cross ratios and have been studied previously in the literature~\cite{Alday:2009zm,Caron-Huot:2021usw,Arkani-Hamed:2023epq,Flieger:2025ekn}. Special subconfigurations of \eqref{three-parameter condition} include the following:

\begin{itemize}

    \item {\bf Configuration $\mathbf{M^2=2m^2}$}: defined by the constraint
    \begin{equation}
        M^2=2m^2 \,,
        \label{double mass condition}
    \end{equation}
    this setup can be obtained by choosing, for $j=1,\dots,4$,
    \begin{equation}
        y_j = m\,e_j \,,
    \end{equation}
    where $e_i$ is the $i$-th standard basis vector in $\mathbb{R}^4$. As we will review in Section \ref{sec:Amplituhedron}, this Coulomb-branch configuration has been studied in~\cite{Arkani-Hamed:2023epq,Flieger:2025ekn}, where it was proposed that its corresponding loop integrands can be obtained from the canonical forms of a particular deformation of the Amplituhedron geometry introduced in \cite{Arkani-Hamed:2013jha,Arkani-Hamed:2013kca}. Moreover, the authors in \cite{Arkani-Hamed:2023epq,Flieger:2025ekn} explicitly computed the scattering amplitude through two loops, 
    showing that the leading IR divergence is governed by the cusp anomalous dimension (whose all-loop expression was derived in \cite{Beisert:2006ez}).

    \item {\bf Configuration $\mathbf{M=0}$}: obtained by setting
    \begin{equation}
        M^2=0\,,
        \qquad
        m^2\neq0 \,,
        \label{single mass condition}
    \end{equation}
    it can be reached by taking
    \begin{equation}
        \bigl(y_1\;y_2\;y_3\;y_4\bigr)
        =
        \frac{m}{2}
        \begin{pmatrix}
            \sqrt{3} & \sqrt{3} & \sqrt{3} & \sqrt{3} \\
            i        & -i       & i        & -i       \\
            \sqrt{2} & 0        & -\sqrt{2}& 0        \\
            0        & \sqrt{2} & 0        & -\sqrt{2}
        \end{pmatrix}.
        \label{M=0 ys config}
    \end{equation}
 This configuration is introduced in~\cite{Alday:2009zm}, where as reviewed above a holographic interpretation is provided for its dual conformal invariance. As in the $M^2=2m^2$ case, the leading IR divergences in this case are given by the cusp anomalous dimension.

    \item {\bf Configuration $\mathbf{m=0}$}: this setup is defined by taking
    \begin{equation}
        m^2=0\,,
        \qquad
        M^2\neq0 \,,
    \end{equation}
    and can be reached by imposing
    \begin{equation}
    y_j
    =
    \frac{M}{\sqrt{2}}\,\mathrm{e}^{i\pi/4}
    \left(
        e_j+i\,e_{j+1}
    \right),
    \label{m=0 ys config}
\end{equation}
    where we take indices modulo $4$. This configuration was introduced in~\cite{Caron-Huot:2021usw}, where a connection with large $R$-charge correlators was established, generalizing the known scattering amplitude/correlation function duality originally found at the origin of the moduli space of ${\cal N}=4$ SYM~\cite{Eden:2010zz}. The corresponding leading IR divergence is governed by the octagon anomalous dimension, computed to all loops in~\cite{Belitsky:2019fan}.
\end{itemize}

Let us note that our choices of $y_i$ vectors for the configurations $M^2=2m^2$ and $M=0$ differ slightly from the representatives used in~\cite{Arkani-Hamed:2023epq,Flieger:2025ekn} and~\cite{Alday:2009zm}, respectively. The latter obey the same mass conditions \eqref{double mass condition} and \eqref{single mass condition}, but do not satisfy the additional orthogonality condition \eqref{y1y3 y2y4 condition} imposed here. We have chosen the convention \eqref{y1y3 y2y4 condition} to obtain a match between the numerator of the integrands in~\eqref{one loop amplitude dual space} and \eqref{two loop amplitude dual space} and the numerator of the canonical forms that will be discussed in the following section, up to possible kinematic-independent constants. Moreover, the use of complex $y_i$ vectors in \eqref{M=0 ys config} and \eqref{m=0 ys config} should be understood as an analytic continuation of the Coulomb-branch kinematics, rather than as a real Coulomb-branch vacuum of ${\cal N}=4$ SYM.

\section{The deformed Amplituhedron}
\label{sec:deformed-Amplituhedron}
%===========================================================

As stated above, one of the goals of the present paper is to generalize the ideas of~\cite{Arkani-Hamed:2023epq,Flieger:2025ekn} to a broader class of Coulomb-branch kinematics. In this section we develop the corresponding geometric construction. We begin by formulating the deformed one-loop geometry in terms of the embedding-space Gram matrix and then analyze how its real positive region changes as the kinematics are varied. This naturally leads to copositivity as the appropriate language for characterizing the allowed regions and their residual boundaries. We then use this structure to construct the two-loop geometry and determine the logarithmic forms compatible with its boundary data. The discussion through two loops will also suggest which ingredients are likely to remain relevant at higher loop order.

%-----------------------------------------------------------
\subsection{Embedding space and Gram coordinates}
\label{sec:embedding-formalism}
%-----------------------------------------------------------

We first fix the embedding-space conventions. External particles and
external data are labelled by $i,j,k,n=1,\ldots,4$ and loop variables by
$\ell,\ell'=1,\ldots,L$, where $L$ is the loop order. 
Let $x_i^\mu$ denote dual coordinates in four dimensions,
\begin{equation}
    p_i^\mu=x_{i+1}^\mu-x_i^\mu \, ,
\end{equation}
and associate an internal mass $m_i$ to each external region. It is useful to identify each pair $(x_i,m_i)$ with a projective vector $X_i \in \bP^{5} $ constructed as
\begin{equation}
    X_i
    =
    \left(
        x_i^\mu,\,
        x_i^2+m_i^2,\,
        1
    \right)
     \,.
    \label{eq:embedding-X}
\end{equation}
Similarly, loop regions will be represented by massless vectors
\begin{equation}
    Y_{\ell}
    =
    \left(
        y_{\ell}^\mu,\,
        y_{\ell}^2,\,
        1
    \right)  \, .
    \label{eq:embedding-Y}
\end{equation}
The variables \(y_\ell\) should not be confused with the \(SO(6)\) vectors \(y_i\) introduced in~\eqref{symmetry breaking pattern}, which specify the Coulomb-branch vacuum and determine the symmetry-breaking pattern.
Then, choosing the embedding-space bilinear form such that
\begin{equation}
    (X_i,X_j)
    =
    (x_i-x_j)^2+m_i^2+m_j^2 \, ,
    \qquad
    (Y_{\ell},X_i)
    =
    (y_{\ell}-x_i)^2+m_i^2 \, ,
    \label{eq:embedding-products}
\end{equation}
and
\begin{equation}
    (Y_{\ell},Y_{\ell'})
    =
    (y_{\ell}-y_{\ell'})^2 \,.
    \label{eq:YY-embedding}
\end{equation}
 By~\eqref{eq:embedding-products} propagators become linear functions of the loop variables in embedding
space.
We should stress that the variables $X_i$ and $Y_{\ell}$ are projective, namely they are defined only under individual rescaling. Moreover, dual conformal transformations of the dual space coordinates crucially become linear transformations in the embedding space.

After passing to split-signature (2,2) in four dimensions, points in the six-dimensional embedding space $\mathbb{R}^{3,3}$ correspond to points in the space of real two-forms $\bigwedge\nolimits^2\bR^4$, where the latter is equipped with the scalar product\footnote{Note that $(X,Y)$ has signature $(2,4)$ while $\langle X,Y\rangle$ has signature $(3,3)$. The two are related by a complex rotation mapping four-dimensional spacetime of signature $(1,3)$ to one of signature $(2,2)$. }
\begin{equation} 
   \langle X,Y\rangle:=X\wedge Y \, .
    \label{eq:bivector-pairing}
\end{equation}
An important advantage of representing the kinematical data as points in $\bigwedge\nolimits^2\bR^4$ lies in the fact that the constraint
\begin{equation}
 \langle Y,Y\rangle=0 \,,
\label{massless constraint}
\end{equation}
allows us to write each loop variable as a wedge product, i.e.
\begin{equation}
    Y=A \wedge B \equiv AB \,,
\end{equation}
for some pair of projective vectors $A$ and $B$. We can think of $Y$ as the line in $\mathbb{P}^3$ passing through the points $A$ and $B$. The massless constraint \eqref{massless constraint} indeed identifies the image of Minkowski space in embedding space with the Pl\"ucker embedding of the Grassmannian ${\rm Gr}(2,4)$ of lines in projective three-space.

A particularly useful parametrization of the external kinematics of a scattering process is provided by the corresponding Gram matrix
\begin{equation}
G_{ij}:=\langle X_i,X_j\rangle \, .
\label{eq:Gram-def}
\end{equation}
According to~\eqref{eq:embedding-products}, the diagonal entries encode the propagator masses,
\begin{equation}
    G_{ii} = \langle X_i,X_i\rangle = 2m_i^2 \, .
\end{equation}
In the following we will be interested in massive configurations, so we will assume that all diagonal entries of the Gram matrix are non-vanishing.
Importantly, the Gram matrix is invariant under dual conformal transformations and is defined up to individual projective rescalings of the embedding-space coordinates $X_i$. Taking into account the latter projective invariance, one can equivalently choose a representative Gram matrix $\mathcal{G}$ such that ${\cal G}_{ii}= 1$.
In terms of the usual kinematic invariants, we can then write
\begin{equation}
{\cal G}_{ij}
=
\frac{x_{ij}^2+m_i^2+m_j^2}{2m_i m_j} \, .
\label{eq:physical-normalized-Gram}
\end{equation}
Note that a given normalized Gram matrix does not determine a unique Coulomb-branch
mass configuration. Indeed, the map from the four-point kinematic data $(s,t,m_i,M_{i})$, where $s=-x_{13}^2$ and $t=-x_{24}^2$ with $x_{ij}=x_i-x_j$,
to the six independent entries of ${\cal G}$ is in general not
injective, so different choices of masses and kinematic invariants can
represent the same point in Gram space. The Gram matrix therefore
encodes the intrinsic dual-conformal data, while a choice of
$(s,t,m_i,M_{i})$ provides a particular physical realization.

For the unit-diagonal integration convention, we write
\begin{equation}
 \mathcal G=\begin{pmatrix}
 1& \mathcal{G}_{12}&\mathcal{G}_{13}&\mathcal{G}_{14}\\
 \mathcal{G}_{12}&1&\mathcal{G}_{23}&\mathcal{G}_{24}\\
 \mathcal{G}_{13}&\mathcal{G}_{23}&1&\mathcal{G}_{34}\\
 \mathcal{G}_{14}&\mathcal{G}_{24}&\mathcal{G}_{34}&1
 \end{pmatrix}.
 \label{eq:general-Gram-coordinates}
\end{equation}
For computational purposes, we will also use a representative of the Gram matrix with non-unit diagonal entries.
For the mass configurations introduced in Section \ref{sec: CB configs of interest} that satisfy the condition \eqref{three-parameter condition} we can identify
\begin{equation}
{\cal G}_{13}
=
1-\frac{s}{2m^2} \, ,
\qquad
{\cal G}_{24}
=
1-\frac{t}{2m^2} \, ,
\qquad
{\cal G}_{i,i+1}
=
1-\frac{M^2}{2m^2} \, .
\end{equation}
It is useful to make a change of variables and define a set of $x,y,z$ parameters as
\begin{equation}
\frac{s}{m^2}
=
-\frac{(1-x)^2}{x} \, ,
\qquad
\frac{t}{m^2}
=
-\frac{(1-y)^2}{y} \, ,
\qquad
\frac{M^2}{m^2}
=
2-\frac{z}{\sqrt{xy}} \, .
\label{eq:xyz-stmM-map}
\end{equation}
Then its embedding in the general Gram coordinates is
\begin{equation}
 \mathcal{G}_{13}=\frac{1+x^2}{2x} \, ,\qquad \mathcal{G}_{24}=\frac{1+y^2}{2y} \, ,\qquad
 \mathcal{G}_{12}=\mathcal{G}_{14}=\mathcal{G}_{23}=\mathcal{G}_{34}=\frac{z}{2\sqrt{xy}} \, .
 \label{eq:three-parameter-Gram-slice}
\end{equation}
In the following we will mostly use the Gram matrix representative given by
\begin{equation}
    G(x,y,z)
    =
    \begin{pmatrix}
        2x & z & 1+x^2 & z\\
        z & 2y & z & 1+y^2\\
        1+x^2 & z & 2x & z\\
        z & 1+y^2 & z & 2y
    \end{pmatrix}.
    \label{eq:G-xyz}
\end{equation}
In particular, we can recover the different configurations presented in Section \ref{sec: CB configs of interest} by taking 
\begin{alignat}{2}
&\textbf{Configuration}\,\mathbf{M^2=2m^2:}
&\qquad &z   = 0 \,, \\
&\textbf{Configuration}\,\mathbf{M=0:}
&\qquad &z   = 2\sqrt{xy} \,, \\
&\textbf{Configuration}\,\mathbf{m=0:}
&\qquad \qquad &x,y  \to0 \,.
\end{alignat}

Let us briefly discuss the symmetries of this configuration in~\eqref{eq:G-xyz}. The Gram matrix in~\eqref{eq:G-xyz} is invariant under $X_1 \leftrightarrow X_3$ and $X_2 \leftrightarrow X_4$, generating the Klein group
\begin{equation}
    K_4\simeq\mathbb Z_2\times\mathbb Z_2 \, .
    \label{eq:K4-symmetry}
\end{equation}
The full dihedral group $D_4$ acts on \eqref{eq:G-xyz} as a crossing symmetry.  In particular, a generator of the rotation subgroup exchanges
\begin{equation}
    s\longleftrightarrow t \qquad \text{or equivalently} \qquad x\longleftrightarrow y \, ,
    \quad z\longmapsto z \, .
    \label{eq:D4-action-kinematics}
\end{equation}

The Euclidean region is determined by
\begin{equation}
    s<0\,, \quad t<0 \qquad \iff \qquad  x>0 \,, \quad y>0 \,.
\end{equation}
For reasons becoming clear in the next section, when studying the geometries associated with the Coulomb-branch configurations of Section \ref{sec: Coulomb branch review} we will instead choose the region
\begin{equation}
    x<0 \, ,
    \qquad
    y<0 \, ,
    \qquad
    (1+x)^2(1+y)^2>4z^2 \, ,
    \qquad
    z^2<4xy \, .
\label{amplit region}
\end{equation}
This distinction already appears in the original deformed-Amplituhedron
construction \cite{Arkani-Hamed:2023epq}, where the deformation parameters are
chosen in the Amplituhedron region and are subsequently analytically
continued to a Euclidean region for integration.

%-----------------------------------------------------------
\subsection{Review of known configurations}
\label{sec:Amplituhedron}
%-----------------------------------------------------------

Before turning to the geometric description of the massive Coulomb-branch amplitudes discussed in Section \ref{sec: Coulomb branch review}, let us briefly review the seminal construction introduced in \cite{Arkani-Hamed:2013jha, Arkani-Hamed:2013kca} to account for the scattering amplitudes in planar ${\cal N}=4$ SYM theory. Here we restrict to a special class of Amplituhedra which is relevant for our discussion. The construction is far more general and applies to arbitrary particle number, helicity sector, and loop order.

In line with the discussion of Section \ref{sec:embedding-formalism}, at the origin of the Coulomb branch the external particles and all internal propagators are massless, so the external vectors $X_i$ satisfy~\eqref{massless constraint}. This implies that the $X_i$ are decomposable and can therefore be written as
\begin{equation}
X_i =  Z_i \wedge Z_{i+1} \equiv (i,i+1) \,,
\label{massless ext kinematics}
\end{equation}
for some four vectors $Z_i$, $i=1,\dots,4$, with $Z_5\equiv -Z_1$. The $Z_i$ are the momentum-twistors commonly used in ${\cal N}=4$ SYM. In this case, the corresponding Gram matrix reads
\begin{equation}
    G = \langle 1234 \rangle \begin{pmatrix}
        0 & 0 & 1 & 0 \\  0 & 0 & 0 & 1 \\ 1 & 0 & 0 & 0 \\  0 & 1 & 0 & 0
    \end{pmatrix} \, ,
    \label{eq:massless_G}
\end{equation}
where the inner product satisfies $\langle 1234 \rangle = \langle 12,34 \rangle=\langle 23,14 \rangle$.

The color-ordered maximally-helicity-violating (MHV) amplitude in planar ${\cal N}=4$ SYM at four points and one loop has only one contribution, given by a massless box Feynman integral. Geometrically, the four-point one-loop MHV Amplituhedron is defined as the region described by the loop lines $AB  \in {\rm Gr}_{\mathbb R}(2,4)$ subjected to the constraints
\begin{align}
 \langle AB,i \,i+1\rangle >0  \, ,
    \quad i=1,\ldots,4 \,, \qquad 
 \langle AB,13\rangle <0 \,,  \qquad 
 \langle AB,24\rangle <0  \,.
 \label{eq:one_loop_ampl}
\end{align}
Similarly, the $L$-loop generalization is given by the space of loop lines $(AB)_\ell$, $\ell=1,\dots,L$, such that each line satisfies the one-loop constraints \eqref{eq:one_loop_ampl} and
\begin{equation}
\langle (AB)_{\ell}, (AB)_{\ell'} \rangle>0 \,, \qquad \forall\,1\leq\ell<\ell'\leq L \,.
\label{eq:multiloop}
\end{equation}
Conjecturally, each such space is a positive geometry, naturally associated with a unique logarithmic top-form $\mathbf{\Omega}^{(L)}$ with logarithmic singularities at all of its boundaries and nowhere else. At $L=1$ this is true, and the corresponding form is given by
\begin{align}
\mathbf{\Omega}^{(1)}(AB;G)= \frac{  \langle X_1, X_3 \rangle  \langle X_2, X_4 \rangle \, \mathrm{d}\mu(AB) }{\langle AB, X_1 \rangle \langle AB, X_2 \rangle \langle AB, X_3\rangle \langle AB, X_4\rangle} \,,
\label{eq:one_loop_form}
\end{align}
with the standard measure on the Grassmannian given by $\mathrm{d}\mu(AB):=\langle \mathrm{d}^2A, AB \rangle \langle \mathrm{d}^2B, AB \rangle$. Eq.~\eqref{eq:one_loop_form} is the integrand of the massless box, which is the same as the integrand of the one-loop four-point MHV amplitude. At two loops, there is no complete proof that the Amplituhedron satisfies the axioms of a positive geometry, and it is in fact known to require a slight extension of this notion to accommodate the presence of \emph{weighted residues}~\cite{Dian:2022tbs}.
Nevertheless, the form is uniquely fixed~\cite{Dian:2024hil} and corresponds to the decomposition of the two-loop integrand in terms of two four-point massless double-boxes, in the $s$ and $t$ channels. Denoting the loop lines by $AB$ and $CD$, the canonical form is given by
\begin{equation}
\begin{aligned}
   \mathbf{\Omega}^{(2)}(AB,CD;G) &= n_{s}(G) \, \left(  \mathbf{\Omega}^{(2)}_{s}(AB,CD;G)  + \mathbf{\Omega}^{(2)}_{s}(CD,AB;G) \right) \\ & \ + n_{t}(G)\,  \left(  \mathbf{\Omega}^{(2)}_{t}(AB,CD;G) + \mathbf{\Omega}^{(2)}_{t} (CD,AB;G)\right) \, ,
\label{eq:form_two_loops}
\end{aligned}
\end{equation}
where we defined the forms corresponding to double-box topologies as
\begin{equation}
    \mathbf{\Omega}^{(2)}_{s} (AB,CD;G) =  \frac{ \langle X_1, X_3 \rangle^2  \langle X_2, X_4 \rangle \, \mathrm{d}\mu(AB) \wedge \mathrm{d}\mu(CD) }{\langle AB, X_1 \rangle \langle AB, X_2 \rangle \langle AB, X_3 \rangle \langle AB,CD \rangle \langle CD, X_1 \rangle \langle CD, X_3 \rangle \langle CD, X_4\rangle} \, ,
    \label{eq:db_form}
\end{equation}
with $\mathbf{\Omega}^{(2)}_{t}$ obtained from it by swapping $\{1,3\} \leftrightarrow \{2,4\}$.
The coefficients $n_{s}(G)$ and $n_{t}(G)$ in~\eqref{eq:form_two_loops} are both fixed to one by requiring the form to have logarithmic singularities on the correct boundaries and unit maximal residues on suitable maximal-codimension boundaries.
More generally, as presented in \cite{Arkani-Hamed:2013jha, Arkani-Hamed:2013kca}, the $L$-loop four-point MHV integrand of planar ${\cal N}=4$ SYM is expected to be equal to~$\mathbf{\Omega}^{(L)}$.

\smallskip

We now review the deformed-Amplituhedron construction of~\cite{Arkani-Hamed:2023epq,Flieger:2025ekn} describing a two-parameter family of Coulomb-branch configurations. This is precisely \textbf{Configuration} $\mathbf{M^2=2m^2}$ listed in Section \ref{sec: CB configs of interest}. The approach followed in the references is to start from the massless kinematic configuration as in~\eqref{massless ext kinematics} and introduce two deformation parameters $x$ and $y$, such that the new embedding points encoding the four-point kinematics are
\begin{align}
    \widehat{X}_1&= (12)+x \, (34) \,, \ & \widehat{X}_2&= (23)+y \,(41) \,, \nonumber \\
    \widehat{X}_3&= (34)+x \, (12) \,, \ & \widehat{X}_4&= (41)+y \, (23) \,.
\label{eq:def_X}
\end{align}
This deformation indeed encodes massive kinematics, as
\begin{equation}
    \langle \widehat{X}_i , \widehat{X}_i \rangle \neq 0 
    \label{eq:massive_Xi}
\end{equation}
for generic $x$ and $y$. In the framework of Section~\ref{sec:embedding-formalism}, the corresponding Gram matrix is given by \eqref{eq:G-xyz} and the maps to kinematics by~\eqref{eq:xyz-stmM-map}, both with $z=0$.

The $L$-loop deformed Amplituhedron geometries considered in~\cite{Arkani-Hamed:2023epq, Flieger:2025ekn} appear for
\begin{equation}
    -1<x,y<0 \, .
    \label{eq:def_ampl_xy_ineq}
\end{equation}
The authors explain that, when imposing the usual loop-Amplituhedron inequalities~\eqref{eq:one_loop_ampl} and~\eqref{eq:multiloop} while keeping the kinematic chamber~\eqref{eq:def_ampl_xy_ineq}, the canonical form of the deformed Amplituhedron at one loop is given by~\eqref{eq:one_loop_form} provided one replaces $X_i$ by $\widehat{X}_i$, up to a kinematic factor in $x$ and $y$. Similarly, at two loops the deformed geometry yields the form in~\eqref{eq:form_two_loops} with unconstrained coefficients $n_s$ and $n_t$. 
These forms reproduce the one- and two-loop Coulomb-branch integrands of \textbf{Configuration} $\mathbf{M^2=2m^2}$ in the field-theory normalization, for which $n_s(G)=n_t(G)=1$, as fixed by generalized unitarity.

\subsection{The one-loop geometry}
\label{sec:one-loop-geometry}
%-----------------------------------------------------------

Having reviewed the original Amplituhedron construction of \cite{Arkani-Hamed:2013jha,Arkani-Hamed:2013kca} and the deformed Amplituhedron of \cite{Arkani-Hamed:2023epq,Flieger:2025ekn}, we now turn to one of the main proposals of our paper, which accounts for the extension of the ideas of \cite{Arkani-Hamed:2023epq,Flieger:2025ekn} to the geometric description of Coulomb-branch configurations encoded by a general Gram matrix as in~\eqref{eq:general-Gram-coordinates}, depending generically on six independent parameters. Instead of starting from a deformation of the massless kinematics as in~\eqref{eq:def_X}, we start directly from the Gram matrix. 

From now on, let us omit the hats from the $\widehat{X}_i$ and let us denote a generic point in embedding space satisfying~\eqref{eq:massive_Xi} simply by $X_i$, for $i=1,\dots,4$. Consider the region in the space of lines $AB \in {\rm Gr}_{\mathbb R}(2,4)$ defined by the constraints
\begin{equation}
    \langle AB \, , X_i\rangle>0 \, ,
    \qquad
    i=1,\ldots,4 \, .
    \label{eq:one-loop-ineq}
\end{equation}
This region has four codimension-one boundaries defined by the equations
\begin{equation}
    \langle AB \, , X_i\rangle=0 \, ,
    \qquad
    i=1,\ldots,4 \, .
    \label{eq:one-loop-boundaries}
\end{equation}
Their simultaneous intersection defines a generalized degree-two Schubert problem and consists generically of two solutions, which we denote by
\begin{equation}
     X_+ \, , \  X_-\in\Gr(2,4) \, .
    \label{eq:Xpm-def}
\end{equation}
The two points \(X_\pm\) have a simple interpretation in embedding space. The four conditions in~\eqref{eq:one-loop-boundaries} cut out a one-dimensional subspace \(\mathbb P^1\) of embedding space. Imposing in addition the embedding-space null condition $\langle Y,Y\rangle=0$
gives a quadratic equation, whose two solutions are precisely \(X_+\) and \(X_-\). The discriminant of this quadratic equation is proportional to the Gram determinant
\begin{equation}
    \det(G) \, ,
\end{equation}
and hence the two solutions coincide on the locus \(\det(G)=0\). Moreover, the two solutions are real if and only if $G$ has signature $(2,2)$.

To study the space cut out by the inequalities~\eqref{eq:one-loop-ineq} it is useful to introduce the following notion. 
A real $n \times n$ matrix $H$ is called \textit{copositive} if
    \begin{equation}\label{eq:copositive-def}
        v^{\top} H v \geq 0 \, , \quad \forall \, v \in \mathbb{R}^n_{\geq 0} \, .
    \end{equation}
Furthermore, it is called \textit{strictly copositive} if the inequality \eqref{eq:copositive-def} is strict for every nonzero vector $v \in \mathbb{R}^n_{\geq 0}$. The collection of all copositive matrices is a proper cone containing the cone of real positive-definite matrices and the cone of symmetric nonnegative matrices. A characterization of copositive matrices is due to Wilfred Kaplan~\cite{KAPLAN2000203}: a real symmetric matrix $H$ is copositive if and only if every principal submatrix of $H$ has no positive eigenvector with associated eigenvalue that is negative.
Several other characterizations are presented in a survey by Ikramov~\cite{Ikramov}.
The problem of deciding whether a matrix is copositive is co-NP-complete, and is an active area of research in optimization. Fortunately, explicit semialgebraic criteria for copositivity for symmetric matrices are known up to dimension $n=4$. For $n=4$, such a criterion was obtained by Li Ping and Feng Yu Yu~\cite{PING1993109}.

Coming back to our problem,
we express a point in embedding space as
\begin{equation}
    Y
    =
    \sum_{i=1}^{4}c_i X_i + d_+ X_+ + d_- X_- \, ,
    \label{eq:AB-expansion}
\end{equation}
for some real coefficients $c_j$ and $d_{\pm}$. We can also set without loss of generality $\langle X_+,X_- \rangle = -1$.
The inequalities in~\eqref{eq:one-loop-ineq} then read
\begin{equation}
    \langle Y ,X_j \rangle = \sum_{i=1}^4 c_i G_{ij} > 0 \, , 
    \label{eq:one-loop-ineq_2}
\end{equation}
for $j=1,\dots,4$, where $c=(c_1,c_2,c_3,c_4)^{\rm T}$, and are subject to the null-condition
\begin{equation}
    \langle Y,Y \rangle = -2 d_+ d_-  + c^{\rm T}Gc = 0 \, .
    \label{eq:YY}
\end{equation}
The inequalities \eqref{eq:one-loop-ineq_2} define a polyhedral cone in
the $c_i$, while~\eqref{eq:YY}
fixes $d_+d_-$ in terms of the quadratic form $c^{\rm T}Gc$. 
In particular, the region splits
into two hyperboloid branches precisely when the quadratic form $c^{\rm T}Gc$ has fixed sign on the cone defined by the $c_i$ parameters. In turn, writing
\begin{equation}
    c^{\rm T}Gc = v^{\rm T}Hv 
\end{equation}
with $H=G^{-1}$ and $v=G c$, this is equivalent to strict copositivity of $G^{-1}$. We therefore established that if $G^{-1}$ is strictly copositive, then the region in ${\rm Gr}_\mathbb{R}(2,4)$ cut out by~\eqref{eq:one-loop-ineq} consists of two disconnected identical sets. A connected component is isolated by imposing a fixed sign condition on either $d_+$ or $d_-$, which is equivalent to fixing the sign of
\begin{equation}
    \langle AB,X_+ \rangle < 0 \, , \qquad \text{or} \qquad \langle AB,X_- \rangle < 0 \, .
    \label{eq:Xpm_ineq}
\end{equation}
The sign of these inequalities makes this construction consistent with the massless case: setting $X_+=(13)$ and $X_-=(24)$, the inequalities~\eqref{eq:one-loop-ineq} and~\eqref{eq:Xpm_ineq} reproduce~\eqref{eq:one_loop_ampl}.

The complex complement of the four hyperplane sections
$\langle AB,X_i\rangle=0$ in ${\rm Gr}(2,4)$ has a one-dimensional space of logarithmic top forms, as follows from the genus and combinatorial-rank results of~\cite{Brown:2025jjg}. A generator is therefore unique up to a function of the external Gram data. A projective invariant form is in fact given by the general box integrand~\eqref{eq:one_loop_form}. However, one of the axioms of a positive geometry requires the form to be such that its highest codimension residues are normalized to $\pm 1$. A direct computation of these maximal residues at $X_\pm$ yields
\begin{equation}
 \operatorname{Res}_{X_\pm} \mathbf{\Omega}^{(1)}
 =\pm\frac{G_{13}G_{24}}{\sqrt{\det G}} \, ,
 \label{eq:one-loop-normalization-dictionary} 
\end{equation}
and hence the canonical form according to the standard definition carries this inverse factor as normalization. In turn, this is the normalization that makes the integral \emph{pure}; therefore, both the geometric construction and the integration perspective indicate a specific normalization. This has to be compared with the kinematic normalization used to express the actual one-loop Coulomb-branch amplitude in terms of the box integral.

The above geometric construction therefore yields a natural extension of the one-loop Amplituhedron geometry introduced in \cite{Arkani-Hamed:2013jha,Arkani-Hamed:2013kca} and later deformed in \cite{Arkani-Hamed:2023epq,Flieger:2025ekn}, and motivates us to set the following definition:
\begin{tcolorbox}[
    colback=gray!8,
    colframe=gray!55,
    boxrule=0.6pt,
    arc=1.5mm,
    title={\bf One-loop deformed Amplituhedron}]
Given a Gram matrix \eqref{eq:G-xyz} with split-signature (2,2) and such that its inverse
$G^{-1}$ is strictly copositive, the \textit{four-point one-loop deformed Amplituhedron}
\begin{equation}
    {\cA}^{(1)}_4(G)
    =
    \overline{
    \left\{
    AB\in\Gr_{\bR}(2,4):
    \langle AB, X_i\rangle>0,\
    \langle AB, X_\pm\rangle<0
    \right\}} 
    \label{eq:Ahat1-definition}
\end{equation}
defines a positive geometry whose canonical form equals the box integrand in~\eqref{eq:one_loop_form}, up to the normalization in \eqref{eq:one-loop-normalization-dictionary}. 
\end{tcolorbox}
The overline in~\eqref{eq:Ahat1-definition} indicates the closure of the open set in $\Gr_{\bR}(2,4)$.
For generic deformation data, the real geometry has $f$-vector equal to \cite{Arkani-Hamed:2023epq}
\begin{equation}\label{eq:f_vect}
    f\big({\cA}^{(1)}_4\big)
    =
    (2,4,6,4,1) \, ,
\end{equation}
in contrast with the $(6,12,10,4,1)$ $f$-vector for the undeformed Amplituhedron, where the latter is a copy of the non-negative
Grassmannian. Geometrically, the deformation resolves the reducible
Schubert intersections of the massless configuration into generic linear
sections. Eq.~\eqref{eq:f_vect} can be verified immediately in the coordinates of~\eqref{eq:AB-expansion}.

%===========================================================
\subsection{The two-loop geometry}
\label{sec:two-loop-geometry}
%===========================================================

Taking into consideration the one-loop deformed geometry introduced in the previous section, we now study its two-loop extension and determine under which conditions its corresponding two-loop canonical form can be decomposed into double-box topologies.

Let us begin with the definition of the two-loop deformed geometry.
\begin{tcolorbox}[
    colback=gray!8,
    colframe=gray!55,
    boxrule=0.6pt,
    arc=1.5mm,
    title={\bf Two-loop deformed Amplituhedron}]
Given a Gram matrix $G$ with split-signature (2,2) and such that its inverse
$G^{-1}$ is strictly copositive, we define the \textit{four-point two-loop deformed  Amplituhedron} as
\begin{equation}
    {\cA}^{(2)}_4(G)
    :=
    \overline{
    \left\{
    (AB,CD)\in
    {\cA}^{(1)}_4(G)
    \times {\cA}^{(1)}_4(G):
    \langle AB,CD\rangle>0
    \right\}} \, .
    \label{eq:Ahat2-definition}
\end{equation}
\end{tcolorbox}
That is, the two-loop geometry is obtained by considering two loop lines $AB$ and $CD$ such that each of them lives in a one-loop geometry ${\cA}^{(1)}_4(G) $ and such that they satisfy the mutual positivity constraint $\langle AB,CD\rangle>0$ as in the undeformed case (see \eqref{eq:multiloop}).

We defined the two-loop geometry at this level of generality, but now we argue that in order for the corresponding canonical form to have a decomposition into planar double boxes the Gram matrix $G$ must satisfy additional constraints. This should be compared to the undeformed case, where the conditions on the Gram matrix $G$ correspond to positivity conditions on the matrix $Z$ whose columns are the momentum-twistors $Z_i$ in \eqref{massless ext kinematics}. In the massless case, copositivity of $G^{-1}$ is compatible with the positivity conditions on $Z$, which for the usual Amplituhedron are that $Z$ must have positive maximal minors. At four points the massless construction is very simple, since the $4 \times 4$ matrix $Z$ has only one maximal minor, namely its determinant $\langle 1234 \rangle$. Therefore, positivity of $Z$ is compatible with copositivity of $G^{-1}$ as one can see from~\eqref{eq:massless_G}. In the massless case, positivity of $Z$ is enough to uniquely determine the correct geometry at all loop orders. This is however not true in the deformed case, where additional conditions are needed at two loops.

Let us now study the boundary structure of the two-loop deformed Amplituhedron.
The key geometric property determining the canonical form is the boundary structure and its interplay with the real region. The codimension-one algebraic boundary of ${\cA}^{(2)}_4(G)$ is supported on the following nine irreducible components
\begin{equation}
    \langle AB, X_i\rangle=0 \, ,
    \qquad
    \langle CD, X_i\rangle=0 \, , 
    \qquad
    \langle AB,CD\rangle=0 \, .
    \label{eq:two-loop-facets}
\end{equation}
Unlike at one loop, the higher-codimension boundary structure depends nontrivially on the deformation parameters, namely the entries of $G$. Many algebraic intersections of
the divisors \eqref{eq:two-loop-facets} fail to meet the defining positivity constraints of ${\cA}^{(2)}_4(G)$, even before imposing additional constraints on $G$. For example, consider the configuration $AB=X_{\pm}$ obtained by imposing the four equations~\eqref{eq:one-loop-boundaries}. We call such locus a \emph{stratum} and regard it as an element of the stratification of the whole boundary in~\eqref{eq:two-loop-facets}. On this stratum
\begin{equation}
    \langle AB,CD \rangle= \langle X_{\pm},CD \rangle<0 \,,
\label{residual example}
\end{equation}
where the last inequality is due to the one-loop conditions on $CD$.
On the other hand,~\eqref{residual example} violates the mutual positivity constraint in~\eqref{eq:Ahat2-definition}. Consequently, the codimension-four boundary where $AB=X_\pm$ does not intersect ${\cA}^{(2)}_4(G)$ in full (real) dimension. We call such a stratum \emph{residual}, and we call \textit{residual arrangement} the union of all residual strata. Strata that are not residual are called \emph{boundaries}. We just showed that the four strata
\begin{equation}
    AB=X_{\pm}\, , \qquad \text{or} \qquad CD=X_{\pm} \, ,
    \label{eq:res_pm}
\end{equation}
are all residual. The same strata are also residual in the usual massless four-point two-loop Amplituhedron~\cite{Dian:2024hil}, where one identifies $X_+=(13)$ and $X_-=(24)$.

The residual arrangement crucially constrains the canonical form. Since by definition such form must have simple poles only on the boundaries of the geometry, once written as a single rational function its numerator should vanish on all components of the residual arrangement. If this were not the case, the form would have a singularity on a stratum which is not a boundary. The numerator of the canonical form is called the \emph{adjoint polynomial}, and its full determination by imposing interpolation with the residual arrangement has been an active research area in the field of positive geometries~\cite{Lam:2022yly,KohnRanestad2020,BruserWeigert2025,Ranestad_2024,Dian:2024hil}. Our next goal is to apply these ideas to the two-loop deformed Amplituhedron, namely to determine under which conditions on $G$ the residual arrangement is such that it fixes the numerator of the canonical form to correctly reproduce the double-box expansion as in~\eqref{eq:form_two_loops}.

To study the constraints on the form $\mathbf{\Omega}^{(2)}(AB,CD;G)$, we start from the ansatz 
\begin{equation}
\begin{aligned}
\frac{ {\cal N}(AB,CD;G) \, \mathrm{d}\mu(AB) \wedge \mathrm{d} \mu(CD)}{ \langle AB,X_1 \rangle \langle AB,X_2 \rangle \langle AB,X_3 \rangle \langle AB,X_4 \rangle \langle AB,CD \rangle \langle CD, X_1 \rangle \langle CD,X_2 \rangle \langle CD,X_3 \rangle \langle CD,X_4 \rangle } \, .
    \label{eq:two-loop-general-form}
    \end{aligned}
\end{equation}
This is the most general form of a rational top-form with at most simple poles on each codimension-one boundary \eqref{eq:two-loop-facets}.
Homogeneity in each loop variable forces the adjoint numerator to be
bilinear in $(AB,CD)$. Writing $X_5=X_+$ and $X_6=X_-$, the numerator can therefore be expressed as
\begin{equation}
    {\cal N}(AB,CD;G)
    =
    \sum_{i,j=1}^{6}
    c_{ij}(G)
        \langle AB,X_i\rangle\langle CD,X_j\rangle \, .
    \label{eq:general-N}
\end{equation}
Permutation symmetry in the loop variables forces $c_{ij}=c_{ji}$.
Thus, before imposing residual conditions, the degrees of freedom of the numerator, up to an overall scale, are 20.

The first constraints arise from the residual strata in~\eqref{eq:res_pm}. One immediately checks that evaluating the numerator~\eqref{eq:general-N} on each such stratum and imposing its vanishing removes from the ansatz all terms involving $X_5$ or $X_6$. That is, for every $i=1,\dots,6$:
\begin{equation}
    c_{5i}=c_{i5}=c_{6i}=c_{i6}=0 \, .
\end{equation}
After this reduction, the degrees of freedom of the numerator reduce to 9.

A direct comparison with the desired decomposition into double-boxes in~\eqref{eq:form_two_loops} shows that all the repeated and adjacent monomials, namely
\begin{equation}
    \langle AB , X_i \rangle \langle CD , X_i \rangle \, , \qquad \text{and} \qquad \langle AB , X_i \rangle \langle CD , X_{i+1} \rangle \, ,
\end{equation}
respectively, for $i=1,\dots,4$ with indices taken modulo 4, should be excluded by the residual arrangement. One verifies that this is achieved if the adjacent strata
\begin{equation}
   \langle AB,X_i\rangle
    =
    \langle AB,X_{i+1}\rangle
    =
    \langle CD,X_i\rangle
    =
    \langle CD,X_{i+1}\rangle
    =0 \,,
    \label{eq:res_adj}
\end{equation}
are all residual. Indeed, evaluating the reduced numerator on~\eqref{eq:res_adj} for $i=1$ leaves the four monomials corresponding to $c_{kl}$ with $k,l \in \{3,4\}$. If the stratum is residual, all such coefficients must therefore vanish. It turns out that in order for the strata in~\eqref{eq:res_adj} to be residual, the Gram matrix must satisfy additional copositivity conditions on sign-twisted minors of its inverse $G^{-1}$. We delegate the details to Appendix~\ref{app: residual arrangement}. Using the known explicit copositivity criteria for $2 \times 2$ matrices, we can translate these conditions into explicit algebraic inequalities on the entries of $G^{-1}$. The stratum in~\eqref{eq:res_adj} is residual if and only if for $\{j,j+1\}=\{1,2,3,4\} \setminus \{i,i+1\}$ we have 
\begin{equation}
    (G^{-1})_{jj}>0 \, ,
    \qquad
    (G^{-1})_{j+1,j+1}>0 \, ,
    \qquad
    |(G^{-1})_{j,j+1}|
    <
    \sqrt{(G^{-1})_{jj}(G^{-1})_{j+1,j+1}} \, .
   \label{eq:2x2-residual-condition}
\end{equation}
Imposing these conditions for every $j=1,\dots,4$ reduces the allowed coefficients of the numerator to $c_{i,i+2}$ for $i=1,\dots,4$. 

We briefly explain the structure of the strata in~\eqref{eq:res_adj} in the massless case. There, the stratum $\langle AB,i-1i\rangle
    =
    \langle AB,ii+1\rangle = 0$ factors into two irreducible components, consisting of lines $AB$ through the point $i$ or those lying on the plane spanned by $i-1,i$ and $i+1$. Therefore,~\eqref{eq:res_adj} factors into four irreducible components; two of them are residual, while the others are boundaries but become residual once intersected with $\langle AB,CD \rangle = 0$~\cite{Dian:2024hil}.

Lastly, we need to make sure that the structures proportional to \(c_{i,i+2}\) are not eliminated by the residual arrangement. For example, the coefficient \(c_{13}=c_{31}\) is isolated by restricting the numerator on the stratum
\begin{equation}
\begin{split}
    \langle AB,X_2\rangle
    &=
    \langle AB,X_3\rangle
    =
    \langle AB,X_4\rangle
    =0 \, ,\\
    \langle CD,X_1\rangle
    &=
    \langle CD,X_2\rangle
    =
    \langle CD,X_4\rangle
    =0 \, .
\end{split}
\label{eq:mixed-nonresidual-example}
\end{equation}
For this structure to survive the residual-interpolation conditions, the corresponding stratum must be non-residual, namely it must contain a genuine boundary component. On this locus the relevant sign-twisted \(2\times2\) principal submatrix of \(G^{-1}\) is
\begin{equation}
    \begin{pmatrix}
(G^{-1})_{11} & -(G^{-1})_{13}\\
-(G^{-1})_{13} & (G^{-1})_{33}
\end{pmatrix}.
\end{equation}
Assuming positive diagonal entries, residuality of this stratum requires strict copositivity of the sign-twisted matrix. Consequently, a sufficient condition for the stratum to be non-residual, and hence to contain a genuine boundary component, is
\begin{equation}
    (G^{-1})_{13}
    >
    \sqrt{(G^{-1})_{11}(G^{-1})_{33}} \, .
    \label{eq:opposite-nonresidual-condition}
\end{equation}
The analogous condition with \((13)\mapsto(24)\) ensures that the structure proportional to \(c_{24}=c_{42}\) is likewise not removed by residual interpolation. The values and non-vanishing normalization of these remaining coefficients are instead fixed by the appropriate canonical boundary residues, as discussed below.

We can summarize this result as follows.
\begin{tcolorbox}[
    colback=gray!8,
    colframe=gray!55,
    boxrule=0.6pt,
    arc=1.5mm,
    title={\bf Two-loop double-box numerator space
    }
]
Assume that the Gram matrix $G$ has split signature $(2,2)$, that $G^{-1}$ is strictly
copositive. Assume further that
$$
    (G^{-1})_{jj}>0 \, ,
    \qquad
    (G^{-1})_{j+1,j+1}>0 \, ,
    \qquad
    |(G^{-1})_{j,j+1}|
    <
    \sqrt{(G^{-1})_{jj}(G^{-1})_{j+1,j+1}} \, ,
$$
for $j \in \{1,2,3,4 \}$, and that
$$
    (G^{-1})_{13}
    >
    \sqrt{(G^{-1})_{11}(G^{-1})_{33}} \,, \qquad     (G^{-1})_{24}
    >
    \sqrt{(G^{-1})_{22}(G^{-1})_{44}} \,.
$$
Then the two-loop numerator~\eqref{eq:general-N} of the canonical form~\eqref{eq:two-loop-general-form} reduces to 
\begin{equation}
\begin{aligned}
    {\cal N}(AB,CD;G)
    =
     \, & n_s(G) \, G_{13}^2G_{24}\left( \langle AB,X_2 \rangle \langle CD,X_4 \rangle + \langle AB,X_4 \rangle \langle CD,X_2 \rangle \right) \\
    +
    &n_t(G) \, G_{24}^2G_{13}\left( \langle AB,X_1 \rangle \langle CD,X_3 \rangle + \langle AB,X_3 \rangle \langle CD,X_1 \rangle \right)  ,
\end{aligned}
    \label{eq:two-loop-double-box-span}
\end{equation}
which yields the double-box expansion in~\eqref{eq:form_two_loops}. At this stage, the coefficients $n_s(G)$ and $n_t(G)$ are unconstrained functions of the external data $G$.
\end{tcolorbox}
Note that the factors of $G_{ij}$ in~\eqref{eq:two-loop-double-box-span} are there to ensure the full form is invariant under independent rescaling of each individual $X_i$.
In the ordinary undeformed four-point two-loop MHV Amplituhedron, further residual strata fix $n_s(G)=n_t(G)$. Their specific numerical value is further fixed by normalizing some maximal-codimension residue. On the other hand, the residual arrangement of the deformed Amplituhedron does not by itself yield this additional constraint. Nevertheless, similarly to the one-loop case, requiring unit maximal residues fixes $n_s(G)$ and $n_t(G)$ up to orientation. For example, the stratum in~\eqref{eq:mixed-nonresidual-example}, seen as a cut configuration, isolates one single double-box topology. Taking a further codimension-two residue we reach the \emph{leading singularity} of the corresponding double-box. A direct computation shows that this leading singularity is equal to $n_t^{\rm LS}(G)^{-1}$, where
\begin{align}
    n_s^{\rm LS}(G)
    ={}
    \frac{\sqrt{
        \det(G)\,
        \bigl(G_{13}^{\,2}-G_{11}G_{33}\bigr)
    }}{G_{13}^2 G_{24}} \, , \qquad 
    n_t^{\rm LS}(G)
    ={}
    \frac{\sqrt{
        \det(G)\,
        \bigl(G_{24}^{\,2}-G_{22}G_{44}\bigr)
    }}{G_{24}^2 G_{13}} \, .
    \label{eq:ct-normalization}
\end{align}
Imposing normalized maximal residues therefore fixes the coefficients up to a sign to:
\begin{equation}
    n_s(G) = n_s^{\rm LS}(G) \, , \qquad n_t(G) = n_t^{\rm LS}(G) \, ,
\end{equation}
which in turn make the double-box integrals \emph{pure}.
Again, it is important to point out that this normalization has to be compared with the explicit coefficients accompanying the double boxes in the expression for the two-loop Coulomb-branch amplitude.

Finally, we show the explicit realization of the conditions \eqref{eq:2x2-residual-condition} and \eqref{eq:opposite-nonresidual-condition} in the three-variable Coulomb-branch configuration $\{x,y,z\}$ described in Section~\ref{sec:embedding-formalism}. The derivation is deferred to Appendix \ref{app: residual arrangement}. The conditions boil down to
\begin{equation}
    x<0 \, ,
    \qquad
    y<0 \, ,
    \qquad
    (1+x)^2(1+y)^2>4z^2 \, ,
    \qquad
    z^2<4xy \, .
    \label{eq:xyz-two-loop-chamber}
\end{equation}
The first inequality involving $z$ enforces the split-signature $(2,2)$ of $G$, whereas the second fixes the residual pattern relevant to the two-loop numerator. Within this chamber in $(x,y,z)$, the inverse Gram matrix $G^{-1}$ is strictly copositive, all adjacent pairwise strata are residual, and the opposite mixed strata supporting the two planar double boxes are nonresidual. Thus the general
residual-arrangement result above applies, and the two-loop numerator is reduced to the double-box span~\eqref{eq:two-loop-double-box-span}. The conditions in~\eqref{eq:xyz-two-loop-chamber} are also consistent with the geometric chamber in $(x,y)$ chosen in~\cite{Arkani-Hamed:2023epq}: for $z=0$ one can consider the connected chamber $-1<x,y<0$.

\section{Analytic results for the three-parameter family}\label{sec:computations}
\subsection{One- and two-loop integrals }
\label{sec:three-parameter-amplitudes}

In the previous section we presented a positive-geometry approach to the computation of the integrands of four-point Coulomb-branch amplitudes whose external kinematics are encoded by a three-parameter Gram matrix. We showed that the one- and two-loop integrands follow the box and double-box topologies of~\eqref{eq:one_loop_form} and \eqref{eq:form_two_loops}, respectively. We now provide an explicit integration of those integrands, following the method of finite differential equations initially introduced in \cite{Caron-Huot:2014lda} and subsequently used in \cite{Flieger:2025ekn}. 

We have defined the one- and two-loop amplitudes in Section \ref{sec: Coulomb branch review}. First, it is important to make contact between the contributions to those amplitudes, the deformed Amplituhedron integrands described in Section \ref{sec:deformed-Amplituhedron}, and the scalar integrals involved in the systems of differential equations (DEs) we will study in the subsequent sections.
At one loop, using the field-theory-normalized form \eqref{eq:one_loop_form} and the mapping \eqref{eq:xyz-stmM-map}, we obtain 
\begin{align}\label{eq:canformint}
   \mathcal{B}^{(1)}(s,t;m,M)= \int_{AB}\mathbf\Omega^{(1)}(AB;G)
=\frac{(1+x^2)(1+y^2)}
 {(1-x)(1-y)\Delta}\,f_5(x,y,z)\, ,
\end{align}
with $G=G(x,y,z)$ as in~\eqref{eq:G-xyz}, where $\mathcal{B}^{(1)}(s,t;m,M)$ is the one-loop box integral presented in \eqref{eq:1loopbox} and evaluated in the kinematic configuration \eqref{three-parameter condition}. The square root $\Delta$ is a function of $x,y$ and $z$ defined in~\eqref{eq:chideltasqrt}. 
The equation above defines the function $f_5$, which is a \emph{pure} box integral of transcendental weight equal to two.
At two loops, the integral of the double-box form in \eqref{eq:db_form} is related to the double-box integral \eqref{B2} as
\begin{align}\label{eq:canformint2}
\mathcal{B}_s^{(2)} (s,t;m,M) = \int_{AB,CD}\mathbf\Omega_s^{(2)}(AB,CD;G)
&=\frac{(1+x^2)^2(1+y^2)}
{(1-x)^2(1+x)(1-y)\Delta}\,
f_{15}(x,y,z) \, ,
\end{align}
and similarly for the $t$-channel, by swapping $s \leftrightarrow t$ in the corresponding normalization.\footnote{The form in \eqref{eq:form_two_loops} comes with a factor $1/2$ in the amplitude, because of the symmetry in the loops.} Eq. \eqref{eq:canformint2} defines $f_{15}$, which is a pure integral of transcendental weight four. 
We should note that we are assuming that at two loops the relative normalizations between the deformed-Amplituhedron canonical forms and the Coulomb-branch integrands are the same as the ones found with generalized unitarity in the $z=0$ case in \cite{Arkani-Hamed:2023epq,Flieger:2025ekn}. The nontrivial IR exponentiation found in Section~\ref{sec:walking} provides a strong consistency check of this assumption.

The important feature of the normalized box and double-box integrals
entering the amplitude, shown in \eqref{eq:canformint} and \eqref{eq:canformint2}, is that they are finite in $D=4$ dimensions and dual conformal invariant (DCI)~\cite{Alday:2009zm,Henn:2011xk}.
Their UV finiteness can be checked by explicit power counting. They are IR finite because Coulomb-branch masses---equivalently, the deformation parameters $(x,y,z)$ ---act as infrared regulators. In particular, on the nondegenerate Euclidean contour considered below, all external propagators are massive (the only internal massless leg in the two-loop box is integrable in the IR). As a consequence, we can readily use the framework developed in~\cite{Caron-Huot:2014lda}, which specializes and simplifies the method of differential equations for finite, DCI integrals in four dimensions. This section is devoted to deriving the analytic results for these one- and two-loop integrals from their differential equations. The DE system is obtained by applying the method presented in~\cite{Caron-Huot:2014lda,Flieger:2025ekn}.

\subsubsection{The box integral}\label{sec:oneloopbox}

In this section we illustrate the differential-equation method through the one-loop box integral, before applying the same framework to the two-loop problem. Massive one-loop box integrals are well known and have been extensively studied in the literature~\cite{tHooft:1978jhc,Bourjaily:2019exo}.

We find it convenient to define a family of one-loop integrals in the embedding space variables introduced in Section \ref{sec:embedding-formalism}:
\begin{equation}
I_{a_1,a_2,a_3,a_4}:=\int_Y\frac{1}
 {(YX_1)^{a_1}(YX_2)^{a_2}(YX_3)^{a_3}(YX_4)^{a_4}}
 \label{one-loop family}
\end{equation}
where the measure is defined as $\int_Y=\int\dd^6Y\,\delta((Y,Y)/2)/
[i\pi^2\operatorname{vol}(GL(1))]$ and DCI requires $a_1+a_2+a_3+a_4=4$. The one-loop box we are interested in is the representative $I_{1,1,1,1}$, which is related to \eqref{eq:1loopbox} simply by 
\begin{equation}
   \mathcal{B}^{(1)}(s,t;m,M)=(1+x^2)(1+y^2) I_{1,1,1,1}(x,y,z)
\end{equation} 

The first step towards the computation of the one-loop box $I_{1,1,1,1}$ by means of the method of finite differential equations consists in writing a set of integration-by-parts (IBP) identities relating representatives of the one-loop family \eqref{one-loop family} with different integer indices $a_i$. Considering that the integrals we are interested in are UV and IR finite, it is convenient to formulate the IBP relations for the family \eqref{one-loop family} directly in embedding space, without making use of dimensional regularization, as exposed thoroughly in \cite{Caron-Huot:2014lda,Flieger:2025ekn}. The present construction is a simple application of the same procedure to our three-parameter problem.

Every member of the one-loop family \eqref{one-loop family} can be written as a linear combination of five basis elements, which we take as \textit{master integrals}. The differential equation takes a remarkably simple form in a basis where all basis elements correspond to $\mathrm{d} \log$ integrands. In such a basis the connection matrix takes a strictly lower-triangular block form, reflecting the organization of the pure master integrals into blocks of uniform \emph{transcendental weight}. 
The integrals~\eqref{one-loop family} generally do not correspond to $\mathrm{d} \log$ integrands, and the actual $\mathrm{d} \log$ basis involves linear combinations of them with algebraic coefficients in $(x,y,z)$. A procedure for finding a $\mathrm{d} \log$ basis is outlined in~\cite{Caron-Huot:2014lda,Flieger:2025ekn}; following it, we obtain the desired master integrals, listed in Appendix \ref{app:oneloopbasis}.

Denoting by $\boldsymbol{f}^{(1)}=(f_1, \dots, f_5)$ the vector of master integrals for the one-loop family \eqref{one-loop family}, the one-loop DE system reads
\begin{equation}\label{eq:canform}
    \mathrm{d} \boldsymbol{f}^{(1)}= \mathrm{d} A^{(1)} \cdot \boldsymbol{f}^{(1)},
    \qquad   A^{(1)}=\sum_{\alpha=1}^{6}R_\alpha \log W_\alpha(x,y,z),
\end{equation}
for constant rational matrices $R_{\alpha}$, where $W_\alpha(x,y,z)$ are the kinematic \textit{letters} of the system.
Explicitly, 
\begin{equation}
A^{(1)}=\begin{pmatrix}
0&0&0&0&0\\[2pt]
2  \log W_1&0&0&0&0\\[2pt]
-\log W_{5}&0&0&0&0\\[2pt]
-\log W_2&0&0&0&0\\[2pt]
0&
-\dfrac{1}{2}\log W_{3}&
-2 \log W_{6}&
\log W_{4}&
0
\end{pmatrix} \,.
\label{eq:one-loop-canonical-system}
\end{equation}
which manifests the three blocks of uniform transcendental weight, with $f_1 \equiv 1$ being weight-zero, $f_2,f_3,f_4$ of weight one, and the box $f_5$ of weight two,
\begin{equation}\label{eq:puref5}
    f_5(x,y,z)=(1-x)(1-y)\Delta\,I_{1,1,1,1} \ .
\end{equation}
Therefore, the one-loop \textit{alphabet} is given by $W_1=x, W_2=y$ and
\begin{align}
 W_{3}&=
\frac{
 P_3+(1+x)(1-y)\Delta
}{
 P_3-(1+x)(1-y)\Delta
},
&
 W_{4}&=
\frac{
 P_4+(1-x)(1+y)\Delta
}{
 P_4-(1-x)(1+y)\Delta
},
\label{one-loop letters 1}
\\
 W_{5}&=
\frac{\chi-z}{\chi+z},
&
 W_{6}&=
\frac{
 (1-x)(1-y)z+\chi\Delta
}{
 (1-x)(1-y)z-\chi\Delta
} \,,
\label{one-loop letters 2}
\end{align}
where  
\begin{align}
&\chi=\sqrt{z^2-4xy},\qquad
 \Delta=\sqrt{(1+x)^2(1+y)^2-4z^2} \label{eq:chideltasqrt} \\
    &P_3=(1+x)^2(1+y^2)-2z^2  , \qquad  P_4=(1+x^2)(1+y)^2-2z^2 \, .
\end{align}
Throughout this section we work in the Euclidean kinematic region, where
\begin{equation}
    0<x<1 \, , \quad 0<y<1 \, , \quad 0<z<2\sqrt{xy} \, . 
\end{equation}
Moreover, we fix the square root branches by \begin{equation}\label{eq:branch-choice}
\chi=i\sqrt{4xy-z^2} \ , \ \ \ \Delta=\sqrt{(1+x)^2(1+y)^2-4z^2}.
\end{equation}
In this region both the one- and two-loop box integrals are analytic and real-valued, making it convenient to integrate and evaluate the DE system. As already pointed out in \cite{Arkani-Hamed:2023epq}, this chamber differs from that used in the geometric derivation in~\eqref{eq:xyz-two-loop-chamber}, where the Gram matrix has the required signature and satisfies the copositivity conditions. 

Crucially, the matrix $A^{(1)}$ in~\eqref{eq:one-loop-canonical-system} is strictly block lower-triangular, and therefore it only couples integrals of successive transcendental weights $w$ and $w+1$. 
This property implies that the formal solution to the system, written as a path-ordered exponential
\begin{equation}\label{eq:pathordexp}
    \boldsymbol{f}^{(1)}(x,y,z) = \mathcal{P} \exp{\int_{\gamma} \mathrm{d} A^{(1)}} \cdot \boldsymbol f^{(1)}  \Big|_{(x_0,y_0,z_0)}
\end{equation}
can be solved exactly given a boundary vector as $\boldsymbol f^{(1)}$ evaluated at a point $(x_0,y_0,z_0)$.
Here the exponential denotes a series of iterated integrals of the connection matrices, which are ordered along the path $\gamma$ by the operator $\mathcal{P}$, as explained below. Because the connection preserves the strict weight grading, any ordered product of $k$ connection matrices evaluated at arbitrary points along the path vanishes when the system has $k$ weight blocks. Hence the path-ordered exponential in~\eqref{eq:pathordexp} truncates after $k-1$ iterated integrations. For the one-loop system, which has three weight blocks of weights $0,1,2$, all products of three connection matrices vanish and the expansion therefore terminates at transcendental weight two.
We should note the difference with the canonical form in the sense of~\cite{Henn:2013pwa}: in that case one works in dimensional regularization, the matrices involved are usually not block triangular, and only the first few orders in the perturbative $\epsilon$-expansion are computed.

The \textit{symbol} $\mathcal S$ \cite{Goncharov:2010jf} of the box-integral $f_5$ can be read off by summing over all paths connecting it to the constant master $f_1$ and yields
\begin{equation}\label{eq:one-loop-symbol}
    \mathcal S\!\left[ f_5 \right]
    =
    -x\otimes W_{3}
    -y\otimes W_{4}
    +2\, W_{5}\otimes W_{6} \,.
\end{equation} 
More importantly, the full solution can then be integrated along any path \(\gamma\) in the three-parameter kinematic space\footnote{We follow the convention in Chen iterated integrals of~\cite{Caron-Huot:2014lda}.
Given one-forms $\omega_r$, and writing $\gamma^*\omega_r=a_r(t)\,\mathrm dt$, define
\[
\int_\gamma\omega_1\cdots\omega_n
 :=
 \int_{0\leq t_1\leq\cdots\leq t_n\leq1}
 a_1(t_1)\cdots a_n(t_n)\,
 \mathrm d t_1\cdots\mathrm dt_n \, .
\]
Thus, $\omega_1$ is integrated nearest to the base point and
$\omega_n$ nearest to the endpoint. For the column-vector system
$\mathrm d \boldsymbol f^{(1)}= \mathrm d A^{(1)}\,\boldsymbol f^{(1)}$, path ordering $\mathcal{P}$ places
matrices evaluated at later times on the left.},
\begin{align}\label{eq:1loop_tree_int}
    f_5(x,y,z)
    =& f_5-\frac{1}{2}f_2\int_\gamma \mathrm{d} \log W_{3}-2f_3\int_\gamma \mathrm{d} \log W_{6}+f_4\int_\gamma \mathrm{d} \log W_{4}\notag\\
 & -\int_\gamma \mathrm{d} \log x\,\mathrm{d} \log W_{3}
    -\int_\gamma \mathrm{d} \log y\,\mathrm{d} \log W_{4}
    +2\int_\gamma \mathrm{d} \log W_{5}\,\mathrm{d} \log W_{6} \,,
\end{align}
where all the master integrals $f_i$ on the right hand side are evaluated at $(x_0,y_0,z_0)$.

There is an obvious candidate boundary point in our setting.
Our Gram matrix in~\eqref{eq:G-xyz} reduces to the configuration studied in \cite{Arkani-Hamed:2023epq,Flieger:2025ekn} when $z\rightarrow0$. Furthermore, the pullback of our system to the $z=0$ slice reduces to the one of~\cite{Flieger:2025ekn}. Therefore, it is most convenient to build on the existing result and choose a one-parameter path to perform the only remaining integration in $z$ from $(x_0,y_0,z_0)=(x,y,0)$, given by
\begin{equation}\label{eq:path}
    \gamma(t)=(x,y,tz) \, ,
    \qquad 0\leq t\leq1 \, .
\end{equation}
Using the boundary data shown in Appendix \ref{app:oneloopbasis} and integrating from $t=0$ to $t=1$ gives
\begin{align}
 f_5(x,y,z)={}&2\log x\log y
 -\log x\log \left[ y^2 W_{3}(x,y,z) \right]
 -\log y\log \left[ x^2 W_{4}(x,y,z) \right] \notag\\
 &-2i\pi\log \left[-  W_{6}(x,y,z) \right] 
  +2\int_0^1
   \log W_5(x,y,tz)\,
   \frac{\mathrm d}{\mathrm dt}\log W_6(x,y,tz)\, dt  \,,
 \label{eq:f5-compact-z}
\end{align}
where in the last term we used $W_5(x,y,0)=1$, and left it unintegrated for compactness. The full analytic expression in terms of Goncharov polylogarithms (GPL) can be obtained by rationalizing the alphabet. We checked the result \eqref{eq:f5-compact-z} numerically against the standard Feynman-parameter representation and obtained agreement to 45 digits at multiple kinematic points in the Euclidean region, as can be reproduced using the ancillary file \texttt{box.m}.
The one-loop amplitude defined in Section~\ref{sec:onetwoloopamp} is then given by
\begin{equation}\label{eq:oneloopamptof5}
    \mathcal M^{(1)}(x,y,z) =
-\frac{(1+x^2)(1+y^2)}
       {(1-x)(1-y)\,\Delta}\,
 f_5(x,y,z) \, .
\end{equation}

\subsubsection{The double-box integral}\label{sec:2loopresult}
Let us now turn to the computation of the two-loop double-box integral \eqref{B2}, which is the main result of this section. 
To that end, we define a double-box family of integrals as
\begin{equation}
 I_{a_1,a_2,a_3,a_4,b_1,b_2,b_3,b_4,c}:=
 \int_{Y_1,Y_2}
 \frac{1}{
 \prod_{i=1}^{4}(Y_1X_i)^{a_i}
 \prod_{i=1}^{4}(Y_2X_i)^{b_i}(Y_1Y_2)^c} \ 
 \label{eq:two-loop-family}
\end{equation}
where DCI now fixes $a_1+a_2+a_3+a_4+c=4$ and $b_1+b_2+b_3+b_4+c=4$.
In this setup the double-box integral $\mathcal{B}_s^{(2)}(s,t,m,M)$ of \eqref{B2} (evaluated in the three-parameter kinematics \eqref{three-parameter condition}) is related to a representative of this family by 
\begin{equation}\label{eq:dbtoamp}
\mathcal{B}_s^{(2)}(s,t;m,M)=(1+x^2)^2(1+y^2)I_{1,1,1,0,1,0,1,1,1} \,.
\end{equation}
The crossed channel follows by the interchange $x \leftrightarrow y$.
The two-loop IBP relations now include contact identities, which relate two-loop integrals to one-loop integrals, and allow the iterative structure seen at one loop to extend to higher-loop systems, as described in~\cite{Caron-Huot:2014lda,Flieger:2025ekn}.
\begin{figure}
    \centering
    \includegraphics[width=0.85\linewidth]{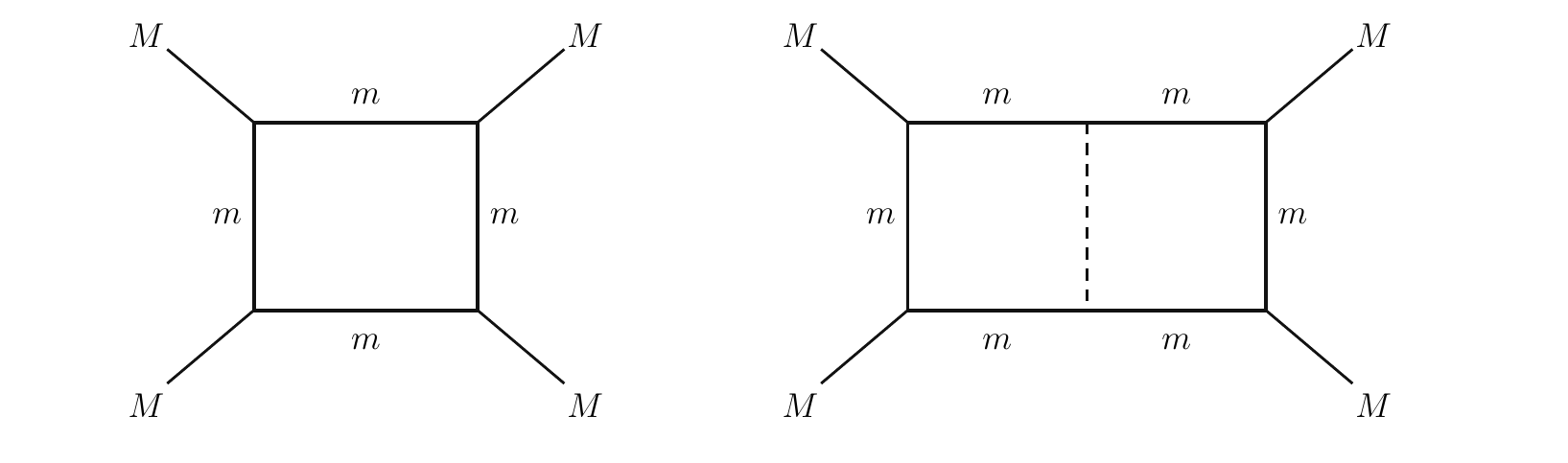}
    \caption{One-loop box (left) and planar two-loop double box (right).
The solid internal propagators have mass $m$, the dashed multiloop one is massless, and the external virtualities are $p_i^2=-M^2$. }
    \label{fig:2loopbox}
\end{figure}
The IBP reduction of the full two-loop family yields 31 master integrals. For the double-box integral of interest, however, the integral itself and all derivatives entering its differential equation close on a 15-dimensional subsystem. This subsystem admits a pure basis of $d\log$ integrals, organized into five blocks of uniform transcendental weights $0,1,2,3,4$ and dimensions $(1,3,7,3,1)$:
\begin{equation}
 \boldsymbol f^{(2)}=
 \bigl(f_1\,\big|\,f_2,f_3,f_4\,\big|\,
f_5,\ldots,f_{11}\,\big|\,
f_{12},f_{13},f_{14}\,\big|\,f_{15}\bigr)^{\mathsf T} \, ,
 \label{eq:two-loop-masters}
\end{equation}
where $f_1,\dots,f_5$ are the one-loop master integrals introduced before and the pure double-box integral corresponds to
\begin{equation}\label{eq:f15norm}
    f_{15}(x,y,z)=(1-x)^2(1+x)(1-y) \, \Delta\, I_{1,1,1,0,1,0,1,1,1}(x,y,z) \, .
\end{equation}

The connection matrix in the sense of~\eqref{eq:canform} has a block-structure with blocks of size $(1,3,7,3,1)$, and we write it as
\begin{equation}
 \mathrm{d} A^{(2)}=
 \begin{pmatrix}
 0&0&0&0&0\\
 \Omega_{10}&0&0&0&0\\
 0&\Omega_{21}&0&0&0\\
 0&0&\Omega_{32}&0&0\\
 0&0&0&\Omega_{43}&0
 \end{pmatrix} \, .
 \label{eq:two-loop-block-system}
\end{equation}
The notation $\Omega_{w,w-1}$ indicates the submatrix connecting the block of transcendental weight $w$ to that of weight $w-1$ by differentiation. The one-loop box system is the closed subsystem formed by the first five components. 
The two-loop alphabet contains twenty letters $\{W_{1},\dots, W_{20}\}$; nine are rational and eleven are genuine algebraic letters, constructed out of the two square roots defined in~\eqref{eq:chideltasqrt}. The one-loop alphabet $\{W_{1},\dots,W_6\}$ is a proper subset of the two-loop one, and this is consistent with our numbering of the letters.
The full alphabet and connection matrix are presented in Appendix~\ref{app:twoloopexpr}, and the basis elements are provided in the ancillary file \texttt{basis.m}.

An important check of our system is that it reduces to the two-loop system of~\cite{Flieger:2025ekn} in the limit $z \rightarrow 0$, and to the system in the massless-external-leg setup of~\cite{Caron-Huot:2014lda} in the limit $z \rightarrow 2\sqrt{xy}$ (see Appendix \ref{app:Boundary data}).

Let us generalize the integration at one loop in~\eqref{eq:1loop_tree_int}, taking 
\(\gamma^*\Omega_{w,w-1}=\Omega_{w,w-1}(t)\, dt\) as the pullback of the connection on the path~\eqref{eq:path}, and denoting with $\boldsymbol f_{[w]}^{(2)}$ the
weight-$w$ block for $w=1,\dots,4$. We can relate the weight-$w$ block to the weight-$(w-1)$ block via
\begin{equation}
 \boldsymbol f_{[w]}^{(2)}(x,y,z)
 =\boldsymbol f_{[w]}^{(2)}(x_0,y_0,z_0)
+\int_{0}^{1}\Omega_{w,w-1}(t)\,
 \boldsymbol f_{[w-1]}^{(2)}(\gamma(t)) \, dt \, .
 \label{eq:weight-recursion}
\end{equation}
The lower-weight vector is obtained by integrating its corresponding DE from \(\gamma(0)\) to \(\gamma(t)\). Hence, iterating this formula produces Chen iterated integrals, defined in matrix form as
\begin{equation}
\mathcal I_\gamma[\Omega_1,\ldots,\Omega_n]
 :=\int_{0\le t_1\le\cdots\le t_n\le1}
 \Omega_n(t_n)\cdots \Omega_1(t_1)\,
 d t_1\cdots d t_n \, .
 \end{equation}
The full three-variable double box result is a function of transcendental weight four expressed as
\begin{align}
 f_{15}(x,y,z)={}&f_{15}
 +\mathcal I_\gamma[\Omega_{43}]\boldsymbol f_{[3]}^{(2)}
 +\mathcal I_\gamma[\Omega_{32},\Omega_{43}]
   \boldsymbol f_{[2]}^{(2)}\notag\\
 &+\mathcal I_\gamma[\Omega_{21},\Omega_{32},\Omega_{43}]
   \boldsymbol f_{[1]}^{(2)}
 +\mathcal I_\gamma[\Omega_{10},\Omega_{21},\Omega_{32},\Omega_{43}]
   f_1 \, ,
 \label{eq:f15-chen-solution}
\end{align}
where the path is chosen as in~\eqref{eq:path} and
each master integral on the right hand side is evaluated at $(x_0,y_0,z_0)=(x,y,0)$. On this boundary point the system decouples into two blocks: one of eleven master integrals corresponding to those in the two-parameter deformed setup of \cite{Flieger:2025ekn}, and the other of four master integrals that can be directly evaluated. We explain this in Appendix \ref{app:Boundary data}. The expressions in terms of classical polylogarithms of the fifteen boundary masters at $z=0$ are supplied in the ancillary file \texttt{boundary.m}.

Rationalizing the square roots $\chi$ and $\Delta$, see Appendix \ref{app:gpltwoloop}, the pullback of \(\mathrm dA^{(2)}\) is a sum of rational \(d\log\) kernels with 26 simple poles. Consequently, the result of \eqref{eq:f15-chen-solution} can be written in terms of a GPL expression, which we store in the ancillary file \texttt{doublebox.m}.
The weight-four symbol of \(f_{15}\) is stored in the ancillary file \texttt{doubleboxsymbol.m} and records the ordered logarithmic kernels in the iterated integrals. It contains all 20 letters \(W_1,\ldots,W_{20}\) and its symbol provides a clean test of the function space the answer belongs to. In particular, to determine whether classical polylogarithms $\operatorname{Li}_n$ suffice, we antisymmetrize the first two and the last two symbol entries separately for each summand in the symbol expression, and then we subtract the expression with the two pairs exchanged, finding that the result cannot be expressed entirely in terms of classical polylogarithms. Equivalently, the weight-four coproduct has a non-vanishing $B_2\wedge B_2$ component, $\delta_{2,2} f_{15}\neq0,$
which is the standard obstruction to representing the result entirely in terms of classical polylogarithms~\cite{Goncharov:2010jf}. This should be compared to the result in the $z\rightarrow 0$ limit in~\cite{Flieger:2025ekn}, which can be expressed in terms of classical polylogarithms.
As for the one-loop result, the answer is real in the Euclidean region, and the corresponding proposed two-loop amplitude becomes

\begin{equation}
     \mathcal M^{(2)}(x,y,z)=
 \frac{(1+x^2)^2(1+y^2)}
 {(1-x)^2(1+x)(1-y)\Delta}\,f_{15}(x,y,z)+\frac{(1+y^2)^2(1+x^2)}
 {(1-y)^2(1+y)(1-x)\Delta}\,f_{15}(y,x,z).
 \label{eq:twoloopamptof15}
\end{equation}
Numerical evaluation of the double-box result agrees with explicit Feynman parameter numerical integration: after integrating the massless leg parameter analytically, the remaining five-dimensional integral can be evaluated with Mathematica
\texttt{NIntegrate}, and can be compared with GPL evaluation with accuracy to 13 digits at various kinematic points in the Euclidean region (see ancillary file \texttt{doublebox.m}).

We mentioned above that the double-box result requires non-classical GPLs. At the same time, its maximal cut defines a rational genus-zero geometry, exhibiting no elliptic obstruction at the level of the maximal cut. This is consistent with the explicit solution of the differential equations, which after rationalization of the two square roots closes within the GPL class. The simplicity of the resulting function space is also compatible with the particularly simple geometry of the four-point embedding-space configuration. Its integrand admits an eight-fold \(d\log\) representation, and the associated finite differential equations can be related directly to this logarithmic structure along the lines of~\cite{HerrmannParraMartinez2020}. The \(d\log\) property alone, however,
does not guarantee a polylogarithmic function space: more general two-loop
Feynman integrals can probe elliptic, higher-genus, and higher-dimensional
algebraic geometries
\cite{Bourjaily:CalabiYauGeometries,Bargiela:TwoLoopGeometries,Cao_2023}.
In particular, higher-point massless double boxes can already be elliptic
\cite{Bourjaily:EllipticDoubleBox,Kristensson:2021ani}, while elliptic
structures also arise on maximal cuts of planar double boxes with internal
masses~\cite{Sogaard:2014jla}. The GPL structure found here should therefore be regarded as a special feature of the four-point massive kinematics considered in this work, at least within the three-parameter family analyzed explicitly here.
It would be interesting to determine whether this polylogarithmic structure persists for the generic six-parameter configurations discussed in Section~\ref{sec:deformed-Amplituhedron}.

\subsection{Interpolating anomalous dimensions}
\label{sec:walking}

The single- and double-box integrals obtained above, valid for the three-scale Coulomb-branch configuration, can readily be used to extract the walking anomalous dimension introduced in~\cite{Alday:2026cpk}. This quantity governs the leading double-logarithmic behavior of the logarithm of the amplitude in a high-energy limit in which the internal Coulomb-branch mass $m$ and the external Euclidean virtuality $M$ are allowed to scale relative to the hard invariants $s$ and $t$.\footnote{By dual conformal invariance, the normalized amplitude depends only on kinematic cross ratios, and hence the limit admits different physical realizations~\cite{Arkani-Hamed:2023epq,Flieger:2025ekn}: increasing $s$ and $t$ at fixed
$m$ and $M$ is equivalent to taking $m,M \rightarrow 0$ at fixed $s$ and $t$. Therefore, the high-energy limit is equivalently a small-mass limit.} Following~\cite{Alday:2026cpk}, we will consider
\begin{equation}\label{eq:walkingparameters}
\omega_s=-\frac{s}{M^2} \, ,\qquad
 \omega_t=-\frac{t}{M^2} \, ,\qquad
 r=\frac{m^2}{M^2} \ .
\end{equation}
We will take $\omega_s,\omega_t\to\infty$ while allowing $r$ to scale so that the \emph{walking parameters}
\[
\eta_s=-\frac{\log r}{\log\omega_s} \, ,\qquad
 \eta_t=-\frac{\log r}{\log\omega_t} \, ,
\]
remain fixed. This limit is implemented using an auxiliary parameter $a$ 
as
\begin{equation}
 \log\omega_s=\kappa_s a \, ,\qquad \log\omega_t=\kappa_t a \, ,\qquad
 \log r=-\rho a \, ,\qquad a\longrightarrow\infty \, ,
 \label{eq:walkingray}
\end{equation}
where $\kappa_s,\kappa_t>0$ and $\rho\in\mathbb R$ are fixed and such that $\eta_s=\frac{\rho}{\kappa_s}, \ \eta_t=\frac{\rho}{\kappa_t}$. The parameter $\rho$ controls the scaling rate of the mass ratio.
Let us note that
$\eta_s\kappa_s=\eta_t\kappa_t=\rho$, and therefore finite physical configurations have either $\eta_s,\eta_t>0$, $\eta_s,\eta_t<0$, or
$(\eta_s,\eta_t)=(0,0)$.

The leading behavior of the logarithm of the normalized amplitude \eqref{eq:cbampdef} defines, in the limit $a \rightarrow \infty$, the walking anomalous dimension as
\begin{equation}\label{eq:walkingdef}
\log\mathcal M
=-\Gamma_{\mathrm{walk}} \, \kappa_s\kappa_t \, a^2+\mathcal{O}(a) \, ,
 \qquad \Gamma_{\mathrm{walk}}=g^2 \Gamma_{\mathrm{walk}}^{(1)}+g^4 \Gamma_{\mathrm{walk}}^{(2)} +O(g^6) \, .
\end{equation}
As the relative scaling of $m$ and $M$ is varied, $\Gamma_{\mathrm{walk}}$ interpolates between two 
regimes.
The $\eta_s=\eta_t=0$ limit is governed by the cusp anomalous dimension $\Gamma_{\rm cusp}$, which controls the universal soft-collinear double logarithms of on-shell massless amplitudes. On the other hand, for $\eta_s=\eta_t=1$ one gets the octagon anomalous dimension $\Gamma_{\rm oct}$, originally identified in the null limit of certain four-point correlation functions of protected operators~\cite{Coronado:2018cxj,Belitsky:2019fan}.

The quadratic growth in $a$ of $\log\mathcal M$ in \eqref{eq:walkingdef} is the familiar Sudakov behavior. At high energies, virtual loop momenta can become simultaneously soft and collinear to the external particles, generating the double logarithms characteristic of this regime~\cite{Sudakov:1954sw,Mueller:1979ih,Collins:1980ih}. These logarithms organize into an exponential form~\cite{Agarwal:2021ais}, such that
\begin{equation}
    \log \mathcal M \sim
-\Gamma_{\rm walk}\log\omega_s\log\omega_t
+\mathcal O(\log\omega_s,\log\omega_t) \, .
\end{equation}
We should note that the $L$-loop contributions ${\cal M}^{(L)}$ to the amplitude generically exhibit a higher-order growth in the IR. Nonetheless, a non-trivial cancellation of IR divergences between the different loop orders of the amplitude yields the quadratic scaling depicted in \eqref{eq:walkingdef} for the logarithm of the amplitude. For example, the quadratic behavior of $\log\mathcal{M}$ at two loops appears as a consequence of a cancellation of higher-order divergences between the two-loop amplitude $\mathcal{M}^{(2)}$ and the square $\bigl(\mathcal{M}^{(1)}\bigr)^2$ of the one-loop amplitude,
\begin{equation}
\log\mathcal{M}=g^2\mathcal{M}^{(1)}
 +g^4\left[\mathcal{M}^{(2)}-\frac12\bigl(\mathcal{M}^{(1)}\bigr)^2\right]+O(g^6) \, .
 \label{eq:logexpansion}
\end{equation}

We will compute $\Gamma_{\text{walk}}$ through two loops by solving the differential equations discussed in Sections \ref{sec:oneloopbox} and \ref{sec:2loopresult} in their IR limit. Such a solution will provide us with the leading IR behavior of the box and double-box integrals \eqref{eq:canformint} and \eqref{eq:canformint2}, which correspondingly determine the leading IR scaling of the one- and two-loop amplitudes ${\cal M}^{(1)}$ and ${\cal M}^{(2)}$. Combining these divergences as in \eqref{eq:logexpansion} we will finally be able to compute the leading IR growth of the logarithm of the amplitude, which as presented in \eqref{eq:walkingdef} is dictated by $\Gamma_{\text{walk}}$.

To set the stage to extract $\Gamma_{\text{walk}}$ from our differential equation systems, let us first parametrize the walking ray~\eqref{eq:walkingray} in terms of our $x,y$ and $z$ variables: combining~\eqref{eq:xyz-stmM-map} and \eqref{eq:walkingparameters}, we get
\begin{equation}
 \omega_s=r\frac{(1-x)^2}{x},\qquad
 \omega_t=r\frac{(1-y)^2}{y},\qquad z
  =\left(2+\frac1r\right)\sqrt{xy} \ .
 \label{eq:walkingsystemmap}
\end{equation}
Recall that in the Coulomb-branch kinematics of the previous sections the external virtualities satisfy $p_i^2=-M_{\rm CB}^2<0$, so that \eqref{eq:xyz-stmM-map} gives $M_{\rm CB}^2/m^2=2-z/\sqrt{xy}$. The walking limit of~\cite{Alday:2026cpk} is obtained after analytically continuing to positive Euclidean virtuality.
In our mostly-plus convention this amounts to $M_{\rm CB}^2\to -M^2$, or equivalently $p_i^2=+M^2>0$, and explains the plus sign in the third relation of~\eqref{eq:walkingsystemmap}.\footnote{The boundary data are fixed at $z=0$ in the Euclidean region $0<z<2\sqrt{xy}$, whereas the walking kinematics lies at $z_{\rm w}>2\sqrt{xy}$. We define the analytic continuation by taking $z$ above the real axis, $z\to z+i0$, and transporting all square roots and logarithms continuously from their branches at $z=0$. In particular, for $z>2\sqrt{xy}$ we have that $\sqrt{4xy-(z+i0)^2}=-i\sqrt{z^2-4xy}$.} As a consequence, 
\begin{equation}
 \begin{aligned}
 x(a)&=\xi\!\left(e^{(\kappa_s+\rho)a}\right) \, ,&
 y(a)&=\xi\!\left(e^{(\kappa_t+\rho)a}\right) \, ,\\
 z(a)&=\left(2+e^{\rho a}\right)\sqrt{x(a)y(a)} \, ,&
 \text{with   } \xi(u)&=\frac{2}{2+u+\sqrt{u(u+4)}} \, .
 \end{aligned}
 \label{eq:exactsystemray}
\end{equation}
On this ray, every letter has the asymptotic form 
\begin{equation}
W_\alpha(x(a),y(a),z(a))=c_\alpha e^{-\nu_\alpha a} \left(1+O(e^{-\epsilon_\alpha a}) \right) \, ,
\label{IR letters}
\end{equation}
where $c_\alpha\neq0$ is independent of $a$, $\nu_\alpha$ is the leading
exponential weight of the letter, and $\epsilon_\alpha>0$ is the positive
gap between the leading and next-to-leading exponential weights of $W_\alpha$. To illustrate the procedure to compute $\Gamma_{\rm walk}$ from the differential equations studied in Sections \ref{sec:oneloopbox} and \ref{sec:2loopresult}, let us now focus on the one-loop case, as the same ideas can then be applied to compute the two-loop contribution to $\Gamma_{\rm walk}$. At one loop the system~\eqref{eq:canform} can be pulled back to the walking ray \eqref{eq:walkingray} as
\begin{equation}
 \frac{d \boldsymbol f^{(1)}}{d a}
  =\bigl(R_{\rm box}+O(e^{-\epsilon a})\bigr)\boldsymbol f^{(1)} \, ,\qquad
 R_{\rm box}=-\sum_{\alpha=1}^{6}\nu_{\alpha}R_{\alpha} \, ,
 \label{eq:residue}
\end{equation}
with $\epsilon:=\min_\alpha\epsilon_\alpha>0$, where the residue matrices $R_\alpha$ are defined in~\eqref{eq:canform}. Thanks to the block triangular structure of the system, the solution is again a truncated series,
\begin{equation}
 \boldsymbol f^{(1)} =e^{aR_{\rm box}} \boldsymbol f^{(1)}_0 +O(e^{-\epsilon' a})
 =\left(1+aR_{\rm box}+\frac{a^2}{2}R_{\rm box}^2\right)\boldsymbol f^{(1)}_0+O(e^{-\epsilon'a}) \ ,
 \label{eq:nilpotenttransport}
\end{equation}
with 
\begin{equation}
\boldsymbol f_0^{(1)}:=\lim_{a\to\infty}e^{-aR_{\rm box}} \boldsymbol f^{(1)} \,,
\label{f0 walking}
\end{equation}
where $0<\epsilon'<\epsilon$ absorbs polynomial factors multiplying the exponentially suppressed corrections. The $\boldsymbol f_0^{(1)}$ vector in \eqref{f0 walking} is an asymptotic boundary vector, not to be confused with the $z=0$ boundary data introduced in previous sections. As discussed in Appendix \ref{app:walking}, the asymptotic matching vector $\boldsymbol f_0^{(1)}$ can be determined by analytically continuing the exact solution from the $z=0$ boundary to the walking ray and expanding it at large $a$. Once $\boldsymbol f_0^{(1)}$ is known, we can extract the IR limit of the master integrals from \eqref{eq:nilpotenttransport}, using the residue matrix $R_{\rm box}$ introduced in \eqref{eq:residue}. The quadratic growth of the one-loop box can then be used to extract the one-loop contribution to the walking anomalous dimension $\Gamma_{\rm walk}$ following \eqref{eq:walkingdef} and \eqref{eq:logexpansion}.

As we will present in the following section, the answer is a piecewise function defined on different chambers, depending on the relative magnitudes of $\kappa_s,\kappa_t,\rho$ or, equivalently, on the values of the two invariant ratios $\eta_s$ and $\eta_t$. The same procedure can be applied to extract the two-loop contribution to $\Gamma_{\rm walk}$, as we present in Section \ref{sec: two-loop gamma walk}.

\subsubsection{One-loop walking dimension}

Let us discuss the computation of the one-loop walking anomalous $\Gamma_{\rm walk}$ dimension in more detail. Since the transcendental weight of the one-loop master integrals is at most two, we have $R_{\text{box}}^3=0$. Using the normalization $f_1=1$ on the $z=0$ slice given in \eqref{eq:one-loop-boundary-data}, and denoting $ \mathbf e_1=(1,0,0,0,0)^{T}$, we obtain 
\begin{equation}
 \Gamma_{\rm walk}^{(1)}
 =-\frac{[a^2](\mathcal M^{(1)})}{\kappa_s\kappa_t} \, ,
 \qquad
 [a^2](\mathcal M^{(1)})
 =-\frac12\bigl(R_{\text{box}}^2 \mathbf e_1 \bigr)_5 \, ,
 \label{eq:one-loop-extraction}
\end{equation}
where $[a^k]$ indicates the coefficient of the $a^k$ power and $(\cdot)_5$ denotes the fifth vector-component.

Since the result is symmetric under \(s\leftrightarrow t\), in the following we restrict the analysis to the region \(\eta_t\leq\eta_s\) (the complementary region is obtained by exchanging the two channels). For \(\rho>0\) we choose \(0<\kappa_s\leq\kappa_t\); when \(\rho<0\) the ordering is reversed, so the same region in \((\eta_s,\eta_t)\) is recovered after interchanging the channel labels. Evaluating the matrix contraction in~\eqref{eq:one-loop-extraction} in the different chambers determined by the relative ordering of \(\rho,\kappa_s\), and \(\kappa_t\) gives the coefficient of the leading \(a^2\) term. Using \(\eta_s=\rho/\kappa_s\) and \(\eta_t=\rho/\kappa_t\) and normalizing by $-\kappa_s\kappa_t$ yields the following one-loop walking anomalous dimension in the \((\eta_s,\eta_t)\) space:
\begin{equation}
\Gamma^{(1)}_{\rm walk}=2\begin{cases}
0 \, ,&\eta_t\leq\eta_s\leq0,\quad \eta_t\leq-1 \, ,\\
(1+\eta_s)(1+\eta_t) \, ,&-1\leq\eta_t\leq\eta_s\leq0 \, ,\\
1+\eta_s+\eta_t-\eta_s\eta_t \, ,&0\leq\eta_t\leq\eta_s\leq1 \, ,\\
1+\eta_s+\dfrac{\eta_t}{2\eta_s}-\dfrac{\eta_s\eta_t}{2} \, ,
 &0\leq\eta_t\leq1\leq\eta_s \, ,\\
1+\dfrac{\eta_t}{2\eta_s}+\dfrac{\eta_s}{2\eta_t} \, ,
 &1\leq\eta_t\leq\eta_s \, .
\end{cases}
\label{eq:G1}
\end{equation}
This exactly agrees with the walking dimension proposed and computed in~\cite{Alday:2026cpk}.
In particular, we recall the all-loop prediction of Eq.~$(4.46)$ in~\cite{Alday:2026cpk}, which reads
\begin{equation}
\Gamma_{\rm walk}(g)=\begin{cases}
0 \, ,
\\
\dfrac{\Gamma_{\rm cusp}(g)}2(1+\eta_s)(1+\eta_t) \, ,
 \\
\dfrac{\Gamma_{\rm cusp}(g)}2+\gamma(g)(\eta_s+\eta_t)
 +\left(\Gamma_{\rm oct}(g)-\dfrac{\Gamma_{\rm cusp}(g)}2-2\gamma(g)\right)
 \eta_s\eta_t \, ,
 \\[4pt]
 (1-\eta_t)\left[
 \widetilde\gamma(g)+
 \left(\dfrac{\Gamma_{\rm cusp}(g)}2+\gamma(g)-\widetilde\gamma(g)\right)\eta_s
 \right]
 +\dfrac{\Gamma_{\rm oct}(g)}4\dfrac{\eta_t}{\eta_s}(1+\eta_s)^2 \, ,
 \\[8pt]
\dfrac{\Gamma_{\rm oct}(g)}2
 \left(1+\dfrac{\eta_t}{2\eta_s}
 +\dfrac{\eta_s}{2\eta_t}\right) \, ,
\end{cases}
\label{eq:walking-all-loop-ansatz}
\end{equation}
where each line is defined in the same domain as \eqref{eq:G1}. The cusp and octagon anomalous dimensions up to two loops are given by
\begin{equation}
 \Gamma_{\text{cusp}}=4g^2-8\zeta_2g^4+O(g^6) \, ,\qquad
 \Gamma_{\text{oct}}=4g^2-16\zeta_2g^4+O(g^6) \, ,
\end{equation}
while the constants $\gamma(g)$ and $\tilde\gamma(g)$ are written in a perturbative expansion as 
\begin{align}\label{eq:walking-interpolating-expansions}
\gamma(g)=\gamma^{(1)}g^2+\gamma^{(2)}g^4+\mathcal O(g^6) \, ,
 \qquad
\widetilde\gamma(g)=\widetilde\gamma^{(1)}g^2
 +\widetilde\gamma^{(2)}g^4+\mathcal O(g^6) \,. 
\end{align}
The latter are free parameters in the conjectural formula and need in principle to be determined order by order.
Comparing \eqref{eq:G1} and \eqref{eq:walking-all-loop-ansatz}, we compute the one-loop values
\begin{equation}
\gamma^{(1)}=\widetilde\gamma^{(1)}=2 \, ,    
\end{equation}
matching the values found in~\cite{Alday:2026cpk} by extracting the double-logarithmic contribution from the dispersive representation of the one-loop box.

\subsubsection{Two-loop walking dimension}
\label{sec: two-loop gamma walk}

We now move to the two-loop analysis, which allows us to determine the previously unknown two-loop constants $\gamma^{(2)}$ and $\widetilde\gamma^{(2)}$ in \eqref{eq:walking-all-loop-ansatz}. The calculation at two loops is completely analogous to the one carried out at one loop in the previous section. Therefore, here we only present the main results, and relegate the details of the computation to Appendix~\ref{app:two-loop walking}. 

The two-loop amplitude ${\cal M}^{(2)}$ scales as $a^4$ in the IR, which is compatible with the transcendental weight four of the analytic answer from Section \ref{sec:2loopresult}. Crucially, the quartic and cubic growth in $a$ cancel when considering the two-loop IR behaviour of the logarithm of the amplitude $\log {\cal M}$ (see \eqref{eq:logexpansion}): this is the two-loop exponentiation of IR divergences.
Following \eqref{eq:walkingdef}, the coefficient of $a^2$ in $\log {\cal M}$ gives the two-loop anomalous dimension  
\begin{equation}
 \Gamma_{\rm walk}^{(2)}
 =-\frac{1}{\kappa_s\kappa_t}
 [a^2]\left[\mathcal M^{(2)}
 -\frac12(\mathcal M^{(1)})^2\right] \, .
 \label{eq:two-loop-extraction}
\end{equation}
Working in the same kinematic region as in the previous section we obtain
\begin{tcolorbox}[
    colback=gray!8,
    colframe=gray!55,
    boxrule=0.6pt,
    arc=1.5mm,
    title={\bf Two-loop walking anomalous dimension
    }
]
\begin{equation}
\Gamma^{(2)}_{\rm walk}=-4\zeta_2\begin{cases}
0 \, ,&\eta_t\leq\eta_s\leq0,\quad\eta_t\leq-1 \, ,\\
(1+\eta_s)(1+\eta_t) \, ,&-1\leq\eta_t\leq\eta_s\leq0 \, ,\\
(1+\eta_s)(1+\eta_t) \, ,&0\leq\eta_t\leq\eta_s\leq1 \, ,\\
1+\eta_s+\eta_t+\dfrac{\eta_t}{\eta_s} \, ,
 &0\leq\eta_t\leq1\leq\eta_s \, ,\\
2+\dfrac{\eta_t}{\eta_s}+\dfrac{\eta_s}{\eta_t} \, ,
 &1\leq\eta_t\leq\eta_s \, ,
\end{cases}
\label{eq:G2}
\end{equation}
\end{tcolorbox}
\noindent
where $\zeta_2=\pi^2/6$ is the second Zeta value.
Comparing our formula with the all-loop conjecture \eqref{eq:walking-all-loop-ansatz}, we can fix the $\gamma^{(2)}$ and $\widetilde\gamma^{(2)}$ coefficients to be
\begin{tcolorbox}[
    colback=gray!8,
    colframe=gray!55,
    boxrule=0.6pt,
    arc=1.5mm,
    title={\bf Determination of two-loop constants from~\cite{Alday:2026cpk}
    }
]
\begin{equation}
 \gamma^{(2)}=\widetilde\gamma^{(2)}= \frac{\Gamma_{{\rm cusp}}^{(2)}}{2} = -4\zeta_2
 \, .
 \label{eq:ourgammas}
\end{equation}
\end{tcolorbox}
\noindent 
We conjecture that this relation between $\gamma$, $\widetilde\gamma$ and $\Gamma_{{\rm cusp}}$ holds to all loop orders. If so, combining it with the ansatz~\eqref{eq:walking-all-loop-ansatz} implies that the walking anomalous dimension $\Gamma_{\rm walk}$ is entirely determined by \(\Gamma_{\rm cusp}\) and \(\Gamma_{\rm oct}\) to all loops.

Furthermore, we can use \eqref{eq:G2} to  study the relationship between the anomalous dimensions $\Gamma_{\rm walk}$ and $\Gamma_{\rm walk}^{\rm FF}$ of the four-point scattering amplitude and the Sudakov form factor (FF), respectively. The latter is the matrix element of a local scalar operator between the vacuum and a two-particle state. Both observables develop a double-logarithmic behavior from soft-collinear regions, which allows us to compare their walking anomalous dimensions. In the near-mass-shell limit studied in \cite{Belitsky:2025bgb} the double logarithms are governed by the octagon anomalous dimension, which corresponds to the ${\eta_s=\eta_t=1}$ endpoint of the walking interpolation. In \cite{Belitsky:2025bgb} the authors showed that the double-logarithmic coefficient of the four-point amplitude and that of the form-factor are related up to three loops as
\begin{equation}
 \left.\Gamma_{\rm walk}^{\rm amp}(\eta,\eta;g)\right|_{\eta=1}
 =2\left.\Gamma_{\rm walk}^{\rm FF}(\eta;g)\right|_{\eta=1}
 =\Gamma_{\rm oct}(g) \, .
 \label{eq:walking-octagon-amplitude-form-factor}
\end{equation}
In~\cite{Alday:2026cpk} the authors showed that the same relation extends to the interior of the walking interpolation at one loop. Comparing our two-loop result in~\eqref{eq:G2} with the two-loop
form-factor anomalous dimension of \cite{Alday:2026cpk} shows that the relation still holds exactly at two-loops on the walking slice $\eta_s=\eta_t=1$.

\section{Conclusions and outlook}
\label{sec:conclusions}

In this work we have studied four-point amplitudes on the Coulomb branch of planar $\mathcal{N}=4$ SYM from a geometric and an analytic perspective. On the geometric side, we extended the deformed-Amplituhedron construction of~\cite{Arkani-Hamed:2023epq,Flieger:2025ekn} to a broader class of massive kinematics, described intrinsically by their embedding-space Gram matrix. The real geometry of the corresponding hyperplane arrangement plays a central role in the construction. In particular, the copositivity properties of the inverse Gram matrix and of appropriate sign-twisted submatrices determine which algebraic intersections can enter the positive region, and hence control the residual arrangement of the geometry.

At one loop, this construction gives us a positive geometric realization of the massive box. At two loops, regularity on the residual arrangement strongly constrains the shape of the logarithmic form: it reduces its numerator to the two-dimensional space spanned by the planar double boxes. This also clarifies the difference between various pieces of information contained in the geometry: the residual arrangement determines the allowed numerator structures, and their relative coefficients and normalization are forced by canonical boundary residues. It would be interesting to understand how this hierarchy of geometric constraints extends at higher loop orders.

On the analytic side, we have worked in the three-parameter Coulomb-branch family relevant for the subsequent infrared analysis. We computed the corresponding double-box integrals in four spacetime dimensions using finite DEs. Using the proposed two-loop double-box decomposition, these analytic expressions determine the corresponding scattering amplitudes and allow us to study their infrared behaviour. In particular, we studied the walking limit introduced in~\cite{Alday:2026cpk}, which interpolates between the cusp and octagon anomalous dimensions. We reproduced the known one-loop result and determined the two-loop double-logarithmic coefficients. The result has the kinematic structure anticipated in~\cite{Alday:2026cpk}, and our result allows us to fix the constants originally left undetermined, showing some unexpected simplifications and testing the conjectural form-factor and amplitude relationship on the two-loop interpolation.

\bigskip
Below we provide a list of possible future directions.

\begin{enumerate}
    \item \textbf{Higher loops.}
An obvious extension of our work is to study higher loop orders. From the amplitude side, even in the most general six-parameter massive Coulomb branch mentioned in this paper or specializations thereof (such as the three-parameter family worked out here), showing that the planar $\mathcal{N}=4$ SYM massive amplitude has the expected decomposition into planar diagrams corresponding to simple deformations of those of massless $\mathcal{N}=4$ SYM requires in principle a computation via Feynman diagrams and generalized unitarity. An alternative is to consider the Britto-Cachazo-Feng-Witten (BCFW) recursion relation~\cite{Britto:2005fq}, as it is discussed in the next research question below. Geometrically, the general strategy is clear within our setup, although the explicit numerator analysis becomes increasingly challenging. A central question is whether the deformed Amplituhedron continues to reproduce, through its canonical form, the correct planar Coulomb-branch integrand for kinematics compatible with the embedding-space formalism. Already at three loops, it would be interesting to determine whether this requires additional constraints on the external Gram matrix beyond those identified here and, if so, whether the resulting conditions eventually stabilize at some finite loop order. At the integrated level, a higher-loop computation of the amplitude would provide a test of the conjectural all-loop relation $\gamma=\tilde{\gamma}=\Gamma_{\rm cusp}/2$ between the cusp anomalous dimension and the undetermined coefficients of the all-loop ansatz proposed in \cite{Alday:2026cpk} for the walking anomalous dimension. Such a relation drastically simplifies the structure of the all-loop ansatz, completely determining it in terms of the cusp and octagon anomalous dimensions.

    \item \textbf{Higher points.}
It is conceptually more challenging to extend the deformed Amplituhedron to higher multiplicity. Such understanding is relevant for example in relation to the BCFW recursion relation~\cite{Britto:2005fq}: four-point higher-loop integrands can be recursively computed by lower-loop higher point ones. There has been evidence for BCFW recursion in certain Coulomb branch configurations at tree level \cite{Herderschee:2019dmc}. Recently, for general recursions involving massive kinematics advancements have been made in~\cite{Bourjaily:2026nlz}. Ideally, the validity of the deformed Amplituhedron construction, if true, would be derived from the BCFW recursion relations, as it was the case for the usual Amplituhedron for massless planar $\mathcal{N}=4$ SYM. Our results suggest a natural starting point for generalizing the geometry to higher points: consider the MHV $n$-point loop Amplituhedron of planar $\mathcal{N}=4$ SYM and replace the special Schubert divisors defining its one-loop boundaries by more general hyperplane sections of $\mathrm{Gr}(2,4)$, in direct analogy with the four-point construction developed here. The nontrivial problem is then to identify the appropriate semialgebraic region in the resulting real hyperplane arrangement whose canonical form reproduces the corresponding Coulomb-branch integrand. At higher loops this construction must in addition be compatible with the mutual-positivity conditions between loop lines. Already at one loop and five points, determining the appropriate deformation and its positive region appears to be an interesting open problem.
    
    \item \textbf{The configuration space of deformed massive kinematics.}
In planar $\mathcal{N}=4$ SYM, the configuration space
${\rm Conf}_n(\mathbb{P}^3)$ of massless external kinematics and its positive
part is intimately connected with the cluster-algebraic and analytic
structures of scattering amplitudes
\cite{Scott:2006,Golden:2013xva,Drummond:2017ssj}.
Here ${\rm Conf}_n(\mathbb{P}^3)$ parametrizes $n$ points in
$\mathbb{P}^3$ modulo projective transformations.
Our results suggest that it would be interesting to develop an analogous
configuration space for massive external kinematics relevant to the Coulomb
branch of $\mathcal{N}=4$ SYM encoded by the deformed Amplituhedron. 
The natural starting point is a configuration of $n$ points in the
projectivized embedding space
$\mathbb{P}(\Lambda^2\mathbb{C}^4)\simeq\mathbb{P}^5$, modulo the induced
projective ${\rm PGL}(4)$ action, with the usual massless kinematics recovered on
the Pl\"ucker quadric $\mathrm{Gr}(2,4)\subset\mathbb{P}^5$. After fixing the
individual projective rescalings, such configurations can be encoded by
normalized symmetric Gram matrices, subjected in particular to the
rank-six constraints of the embedding space.
At four points, our analysis indicates that the real positive region relevant
for Coulomb-branch amplitudes is naturally controlled by copositivity
properties of the inverse Gram matrix and of appropriate sign-twisted
principal submatrices. It would be interesting to understand the
higher-point generalization of these conditions and whether the resulting
positive configuration space possesses structures analogous to the cluster
algebra of ${\rm Conf}_n(\mathbb{P}^3)$. The two-loop analytic data presented
in this paper provide a simple testing ground for possible massive
generalizations of the structures governing singularities, symbol alphabets,
and adjacency properties of planar $\mathcal{N}=4$ SYM amplitudes.

\item \textbf{Positive geometry and integration.} Finally, it would be valuable to connect the geometric and analytic sides of the construction more directly. In particular, the finite differential equations used here should ultimately admit an interpretation in terms of logarithmic integrands, their cuts and boundary geometry, along the lines of~\cite{Caron-Huot:2014lda,HerrmannParraMartinez2020}. Establishing such a dictionary could provide a recursive route from the geometry of the deformed Amplituhedron to the analytic structure of Coulomb-branch amplitudes at higher loop order. On the other hand, deformed Amplituhedra serve as one of the ideal testing grounds for exploring the role of positive geometries in the structure of differential equations of finite integrals.

\end{enumerate}

\acknowledgments
We thank Benjamin Basso, Wojciech Flieger, Johannes Henn, Jungwon Lim, {Antonela Matija\v{s}i\'c}, Rourou Ma, Anders Schreiber, Jaroslav Trnka and Qinglin Yang
for helpful discussions. We also thank Luis F. Alday, Elisabetta Armanini, Andrei V. Belitsky, Kelian H\"aring and Alexander
Zhiboedov for sharing their updated results on the Sudakov anomalous dimension, and for useful comments on the draft. E.M. thanks the Institute for Advanced Study and an anonymous donor for
support during the 2026-27 academic year. F.C. is supported by the
Department of Energy (Grant No. DE-SC0011632), the Walter Burke Institute for Theoretical Physics, and the
Leinweber Forum for Theoretical Physics. M.L. is supported by UKRI Future Leaders Fellowship ``The materials approach to quantum spacetime'' under reference MR/X034453/1. S.H. acknowledges financial
support from University of Hyderabad under the startup
grant (OH-35-CA-2026-27).
S.H. and E.M. acknowledge support from the European Union through the European
Research Council Synergy Grant UNIVERSE+ (no.\ 101118787).
{\tiny
Views and opinions expressed are, however, those of the authors only and do
not necessarily reflect those of the European Union or the European Research
Council Executive Agency. Neither the European Union nor the granting
authority can be held responsible for them.
}

\paragraph{Use of AI-assisted tools.}
The authors used ChatGPT (OpenAI) and Claude Code (Anthropic) for coding assistance, language editing and consistency checks. All suggestions were reviewed by the authors, who take full responsibility for the content of the manuscript.

\break 

\appendix
\label{app:A}
\section{Derivation of the two-loop residual arrangement}
\label{app: residual arrangement}

We present here further details and derivations relevant for the discussion in Section \ref{sec:two-loop-geometry} on the geometry of the two-loop deformed Amplituhedron. We showed how the residual arrangement of the two-loop positive geometry ${\cA}^{(2)}_4(G)$ could be used to constrain the corresponding two-loop canonical form. We will provide here a detailed derivation of the residual arrangement.

Before proceeding with the derivation, let us give a precise definition of the residual arrangement. Starting from the irreducible components of the algebraic boundary, let us
consider the irreducible components of their successive intersections.
Given such a complex stratum $S$, with real part $S(\mathbb{R})$, we define
\begin{equation}
    S_{\geq0}
    :=
    S(\bR)\cap{\cA}^{(2)}_4(G) \, .
\end{equation}
We call $S$ a \emph{boundary} when $S_{\geq0}$ contains a real
component of the expected (maximal) dimension, and \emph{residual} otherwise. The
union of all residual strata defines then the \emph{residual arrangement} ${\cal R}^{(2)}_4(G)$.
Regularity of a logarithmic form requires the residual arrangement to be contained in the vanishing locus ${\cal V}$ of its numerator ${\cal N}$, 
\begin{equation}
    {\cal R}^{(2)}_4(G)
    \subset
    {\cal V}({\cal N}) \, ,
    \label{eq:adjoint-interpolation-app}
\end{equation}
so the residual arrangement gives interpolation conditions on the
numerator. This is the analogue, for the present loop geometry, of the adjoint
interpolation problem familiar from positive geometries
\cite{Arkani-Hamed:pos_geom,KohnRanestad2020,Ranestad_2024,Dian:2024hil}:
one asks to what extent the canonical numerator is characterized by requiring
its vanishing locus to interpolate the residual arrangement. For polytopes,
tree Amplituhedra, and particular loop Amplituhedron geometries, such
interpolation conditions can be sufficient to determine the adjoint, and hence
the numerator of the canonical form, up to an overall normalization
\cite{KohnRanestad2020,Ranestad_2024,Dian:2024hil}. At two loops in the
deformed geometry considered here, they instead determine only the allowed
numerator space, while the remaining normalization data are fixed by canonical
boundary residues. This distinction will be important below.

The dependence on the external data can be expressed directly in terms of
the inverse Gram matrix. In the basis $X_1, X_2, X_3, X_4, X_+, X_- ,$
with $\langle X_+,X_-\rangle=-1$, one has
\begin{equation}
    \langle AB,CD\rangle
    =
    \sum_{i,j=1}^{4}
    (G^{-1})_{ij}
    \langle AB,X_i\rangle
    \langle CD,X_j\rangle
    -
    \langle AB,X_+\rangle
    \langle CD,X_-\rangle
    -
    \langle AB,X_-\rangle
    \langle CD,X_+\rangle \, .
    \label{eq:mutual-bracket-G-app}
\end{equation}
We can fix the projective scaling of $AB$ and $CD$ such that
\begin{equation}
    \langle AB,X_+\rangle
    =
    \langle CD,X_+\rangle \, .
\end{equation}
Using the Pl\"ucker equation $\langle AB,AB\rangle=0$ and similarly for $CD$,
while defining
\begin{equation}
    v_i
    :=
    \langle AB,X_i\rangle-\langle CD,X_i\rangle \, ,
    \qquad i=1,\ldots,4 \, ,
    \label{eq:v-coordinates-app}
\end{equation}
we can rewrite~\eqref{eq:mutual-bracket-G-app} as
\begin{equation}
    \langle AB,CD\rangle
    =
    -\frac12\,v^{\rm T}G^{-1}v \, .
    \label{eq:quadratic-mutual-bracket-app}
\end{equation}

On a product-boundary stratum of $\mathcal{A}^{(1)}_4(G) \times \mathcal{A}^{(1)}_4(G)$ by definition of the one-loop geometry in~\eqref{eq:Ahat1-definition} each $v_i$ is constrained to be positive,
negative, zero, or unrestricted. The stratum therefore determines a union
of orthants $C(S)\subset\bR^4 $.
More explicitly, let
$\Sigma(S)\subset\{0,\pm1\}^4$ denote the set of sign patterns compatible
with $S$. For
$\delta=(\delta_1,\ldots,\delta_4)\in\Sigma(S)$, let us define the support set $I_\delta$ and matrix $D_\delta$ as
\begin{equation}
    I_\delta:=\{i:\delta_i\neq0\} \, ,
    \qquad
    D_\delta
    :=
    \operatorname{diag}(\delta_i)_{i\in I_\delta} \, .
\end{equation}
The corresponding orthant consists of vectors with
$v_i=\delta_i u_i$, $u_i>0$, for $i\in I_\delta$, and $v_i=0$ otherwise. Since mutual positivity requires $\langle AB,CD\rangle>0$,
the stratum is residual when
\begin{equation}
    v^{\rm T}G^{-1}v>0 \, ,
    \qquad
    \forall\,v\in C(S)\setminus\{0\} \, .
    \label{eq:residual-main-app}
\end{equation}
For a fixed orthant write $v_{I_\delta}=D_\delta u$ with
$u\in\bR_{\geq0}^{I_\delta}$.  Residuality on that orthant is therefore
equivalent to strict copositivity of the sign-twisted principal submatrix
\begin{equation}
    H_\delta
    :=
    D_\delta
    (G^{-1})_{I_\delta,I_\delta}
    D_\delta \, ,
    \label{eq:twisted-minor-app}
\end{equation}
where $(G^{-1})_{I_\delta,I_\delta}$ denotes the minor of $G^{-1}$ to the indices $I_\delta$.
If the allowed cone is a union of orthants, every corresponding
$H_\delta$ must be strictly copositive, i.e.
\begin{equation}
H_{\delta} {\rm \; is \; strictly \; copositive} \; \forall \delta \in \Sigma(S) \quad \iff \quad S_{\geq 0} \subset {\cal R}^{(2)}_4(G) \,.
\end{equation}
Thus, $G^{-1}$ controls the one-loop real
chamber while its sign-twisted principal submatrices control the two-loop
residual arrangement.

As a first example, on the stratum $AB=X_+$ one has
$v_i=-\langle CD,X_i\rangle$, and therefore
\begin{equation}
    \langle X_+,CD\rangle
    =
    -\frac12\,v^{\rm T}G^{-1}v<0 \, ,
\end{equation}
if $G^{-1}$ is strictly copositive. Consequently, this locus is incompatible with mutual positivity and it is residual.  The
same argument applies to
\begin{equation}
    AB=X_\pm \, ,
    \qquad
    CD=X_\pm \, .
    \label{eq:Xpm-residual-app}
\end{equation}

A second simple class of residual strata follows from the same argument. Fix any \(i \in \{1,\ldots,4 \}\) and the codimension-six stratum where
\begin{equation}
    \langle AB,X_j\rangle
    =
    \langle CD,X_j\rangle
    =0\,,
    \qquad \forall \,  j \in \{1,2,3,4\} \setminus \{i\} \,.
    \label{eq:matching-triple-residual}
\end{equation}
On this locus the vector \(v\) entering~\eqref{eq:quadratic-mutual-bracket-app} has support on a single coordinate, so that
\begin{equation}
    v^{\rm T}G^{-1}v
    =
    (G^{-1})_{ii}\,v_i^2\,.
\end{equation}
Strict copositivity of \(G^{-1}\) implies in particular\footnote{One applies simply the definition of copositivity to the $i$-th standard basis vector.} \((G^{-1})_{ii}>0\), and hence the sign of the mutual bracket is incompatible with mutual positivity for every generic point on~\eqref{eq:matching-triple-residual}. Since this is true for every $i=1,\dots,4$, the corresponding four strata are therefore residual simply from copositivity of \(G^{-1}\). This observation has an immediate consequence for the numerator in~\eqref{eq:general-N}. After the residual conditions~\eqref{eq:Xpm-residual-app} have removed all terms involving \(X_\pm\), restricting the remaining ten-parameter ansatz to~\eqref{eq:matching-triple-residual} leaves only
\begin{equation}
    \left.{\cal N}\right|_{\eqref{eq:matching-triple-residual}}
    =
    c_{ii}\,
    \langle AB,X_i\rangle
    \langle CD,X_i\rangle\,.
\end{equation}
Regularity of the canonical form therefore requires \(c_{ii}=0\) for \(i=1,\ldots,4\), reducing the numerator from ten parameters to the six off-diagonal bilinears $\langle AB,X_i\rangle
    \langle CD,X_j\rangle$. Further reduction of this six-dimensional space depends on genuinely non-trivial copositivity conditions for sign-twisted principal submatrices of \(G^{-1}\), to which we now turn.

Let us consider a pair of indices
$\{i,j\}\subset\{1,2,3,4\}$ and the product-boundary stratum
\begin{equation}
    \langle AB,X_i\rangle
    =
    \langle AB,X_j\rangle
    =
    \langle CD,X_i\rangle
    =
    \langle CD,X_j\rangle
    =0 \, .
    \label{eq:double-boundary-stratum-app}
\end{equation}
Let $\{k,l\}
    =
    \{1,2,3,4\}\setminus\{i,j\}$
be the complementary pair. On this stratum one has
$v_i=v_j=0$, while $(v_k,v_l)$ can take either sign independently.
Consequently, the allowed cone $C(S)$ is the union of the four quadrants
in the $(v_k,v_l)$-plane.
Residuality requires copositivity of $G^{-1}$
on each of these quadrants. Equivalently, every sign-twist of the
$2\times2$ principal submatrix
\begin{equation}
    H_{kl}
    :=
    \begin{pmatrix}
        (G^{-1})_{kk} & (G^{-1})_{kl}\\
        (G^{-1})_{kl} & (G^{-1})_{ll}
    \end{pmatrix}
\end{equation}
must be strictly copositive. The matrix
$
    \begin{pmatrix}
        a&c\\ c&b
    \end{pmatrix}
$
is strictly copositive if and only if~\cite{PING1993109}
\begin{equation}
    a>0 \, ,\qquad b>0 \, ,\qquad
    c+\sqrt{ab}>0 \, ,
\end{equation}
and hence we must impose this criterion both on $H_{kl}$ and on its nontrivial
sign-twist, obtained by $(G^{-1})_{kl}\rightarrow-(G^{-1})_{kl}$.
The two conditions therefore combine to give
\begin{equation}
    (G^{-1})_{kk}>0 \, ,
    \qquad
    (G^{-1})_{ll}>0 \, ,
    \qquad
    |(G^{-1})_{kl}|
    <
    \sqrt{(G^{-1})_{kk}(G^{-1})_{ll}} \, .
    \label{eq:2x2-residual-condition-app}
\end{equation}

On the contrary, as discussed in Section~\ref{sec:two-loop-geometry}, for
the numerator to yield an expansion in planar double boxes one needs the
two-dimensional stratum
\begin{equation}
\begin{split}
    \langle AB,X_2\rangle
    &=
    \langle AB,X_3\rangle
    =
    \langle AB,X_4\rangle
    =0 \, ,\\
    \langle CD,X_1\rangle
    &=
    \langle CD,X_2\rangle
    =
    \langle CD,X_4\rangle
    =0 \, ,
\end{split}
\label{eq:mixed-nonresidual-example-app}
\end{equation}
to remain a genuine boundary. Its allowed cone is characterized by
$v_1>0$ and $v_3<0$. The corresponding sign-twisted $2\times2$ principal
submatrix of $G^{-1}$ is therefore
\begin{equation}
    \begin{pmatrix}
        (G^{-1})_{11} & -(G^{-1})_{13}\\
        -(G^{-1})_{13} & (G^{-1})_{33}
    \end{pmatrix}.
\end{equation}
The stratum is residual precisely when this matrix is strictly copositive.
Assuming $(G^{-1})_{ii}>0$ for $i=1,3$, the $2\times2$ copositivity
criterion discussed above shows that strict copositivity holds provided that
\begin{equation}
    (G^{-1})_{13}
    <
    \sqrt{(G^{-1})_{11}(G^{-1})_{33}} \, .
\end{equation}
Consequently, a sufficient condition for the stratum
\eqref{eq:mixed-nonresidual-example-app} to remain non-residual, and hence
to be a genuine boundary, is
\begin{equation}
    (G^{-1})_{13}
    >
    \sqrt{(G^{-1})_{11}(G^{-1})_{33}} \, .
    \label{eq:opposite-nonresidual-condition-app}
\end{equation}
The analogous condition with $(1,3)$ replaced by $(2,4)$ gives the second
condition required for the planar double-box topology.

We now specialize the above general conditions to the three-parameter family
\eqref{eq:G-xyz}. We write
\begin{equation}
    \det(G)
    =
    (1-x)^2(1-y)^2\Delta^2 \, ,
\end{equation}
where according to the notation in~\eqref{eq:chideltasqrt} we have
\begin{equation}
    \Delta^2
    =
    (1+x)^2(1+y)^2-4z^2 \, .
    \label{eq:Delta-xyz-residual-app}
\end{equation}
Together with the fact that the two $K_4$-odd eigenvalues of $G$ are $-(1-x)^2$ and $-(1-y)^2$, one has,
away from the degenerate loci $x=1$ and $y=1$,
\begin{equation}
    \operatorname{sgn}(G)=(2,2)
    \qquad\Longleftrightarrow\qquad
     (1+x)^2(1+y)^2-4z^2  >0 \, .
\end{equation}

For the inverse Gram matrix set
\begin{equation}
    a:=(G^{-1})_{11}=(G^{-1})_{33} \, ,
    \qquad
    b:=(G^{-1})_{22}=(G^{-1})_{44} \, ,
    \qquad
    c:=(G^{-1})_{12} \, .
\end{equation}
Then
\begin{align}
    a
    =
    \frac{2\bigl[z^2-x(1+y)^2\bigr]}
    {(1-x)^2\Delta^2} \, , \qquad
    b
    =
    \frac{2\bigl[z^2-y(1+x)^2\bigr]}
    {(1-y)^2\Delta^2} \, , \qquad
    c
    =
    -\frac{z}{\Delta^2} \, ,
    \label{eq:Ginv-xyz-abc-app}
\end{align}
while the opposite entries are
\begin{align}
    (G^{-1})_{13}
    =
    a+\frac{1}{(1-x)^2} \, , \qquad
    (G^{-1})_{24}
    =
    b+\frac{1}{(1-y)^2} \, .
    \label{eq:Ginv-xyz-opposite-app}
\end{align}
For
\begin{equation}
    x<0 \, ,
    \qquad
    y<0 \, ,
    \qquad
    (1+x)^2(1+y)^2-4z^2 >0 \, ,
\end{equation}
the diagonal entries $a,b$ are automatically positive. Moreover, since
\begin{equation}
    ab-c^2
    =
    \frac{4xy-z^2}
    {(1-x)^2(1-y)^2\Delta^2} \, ,
    \label{eq:adjacent-minor-factorization-app}
\end{equation}
by~\eqref{eq:2x2-residual-condition-app} every adjacent stratum~\eqref{eq:double-boundary-stratum-app} with $j=i+1$ is residual precisely when
\begin{equation}
    z^2<4xy \, .
    \label{eq:adjacent-residual-wall-app}
\end{equation}
The same inequality also implies strict copositivity of the full
$4\times4$ matrix $G^{-1}$.  Indeed, for $u_i\geq0$ define
\begin{equation}
    U:=u_1+u_3,
    \qquad
    V:=u_2+u_4.
\end{equation}
Using~\eqref{eq:Ginv-xyz-opposite-app} we can write
\begin{align}
    u^{\rm T}G^{-1}u
    =
    aU^2+bV^2+2cUV
    +\frac{2u_1u_3}{(1-x)^2}
    +\frac{2u_2u_4}{(1-y)^2} \, .
    \label{eq:Ginv-copositive-proof-app}
\end{align}
If $c\geq0$ this is manifestly positive for nonzero $u\geq0$; if $c<0$,
strict copositivity follows from $c^2<ab$, which by
\eqref{eq:adjacent-minor-factorization-app} is again equivalent to
$z^2<4xy$.

Finally, the opposite mixed strata~\eqref{eq:mixed-nonresidual-example-app} are automatically nonresidual.  Indeed,
\begin{equation}
    (G^{-1})_{13}-(G^{-1})_{11}
    =
    \frac{1}{(1-x)^2}>0 \, ,
    \qquad
    (G^{-1})_{24}-(G^{-1})_{22}
    =
    \frac{1}{(1-y)^2}>0 \, ,
\end{equation}
guaranteeing the desired inequality~\eqref{eq:opposite-nonresidual-condition-app}.

Consequently, all conditions entering the boxed general result reduce for
the family~\eqref{eq:G-xyz} to the open chamber~\eqref{eq:xyz-two-loop-chamber}.

\section{One-loop pure basis}\label{app:oneloopbasis}
In this appendix we spell the one-loop system master integrals explicitly.
The weight-zero and weight-one masters used in Section~\ref{sec:oneloopbox} are
\begin{align}
 f_1={}&6y^2I_{0,0,0,4},\\
 f_2={}&\frac{3y^2(x^4-1)}{x^2}I_{0,0,0,4}
 +\frac{(1-x^2)^3}{x}I_{1,0,3,0},\\
 f_3={}&\chi\left[
 \frac{3yz(1+x^2)}{2x^2}I_{0,0,0,4}
 -\frac{z(1-x^2)^2}{2xy}I_{1,0,3,0}
 +\frac{P_x}{y}I_{1,1,2,0}\right],\\
 f_4={}&\frac{1-y}{1+y}\left[
 \frac{3yz^2(1+x^2)}{2x^2}I_{0,0,0,4}
 -\frac{z^2(1-x^2)^2}{2xy}I_{1,0,3,0}
 +\frac{zP_x}{y}I_{1,1,2,0}
 +P_yI_{0,1,2,1}\right],
 \label{eq:one-loop-lower-weight-basis}
\end{align}
where
\begin{equation}
 P_x=y(1+x)^2-z^2,
 \qquad
 P_y=x(1+y)^2-z^2.
\end{equation}
Together with the box master \(f_5\) in~\eqref{eq:puref5}, these definitions give the ordered
basis used in~\eqref{eq:one-loop-canonical-system}. 
The corresponding values of the masters on the $z=0$ boundary slice are
\begin{align}\label{eq:one-loop-boundary-data}
(f_1, f_2,f_3,f_4,f_5) \Big|_{(x,y,0)}
   &=(1,2\log x,i\pi,-\log y,2\log x\log y) \, , \nonumber \\
 ( W_{3}, W_{4}, W_{5}, W_{6}) \Big|_{(x,y,0)}
   &=(y^{-2},x^{-2},1,-1) \, .
\end{align}

\section{Details on the two-loop computation}\label{app:twoloopexpr}
This appendix supplies the ingredients needed to reproduce the two-loop
double-box result. We first give the alphabet and the complete canonical
differential system, then determine its boundary data, and finally
describe the rationalization and GPL representation used in the ancillary
file \texttt{doublebox.m}.

\subsection{Differential system}
This subsection specifies the 20-letter alphabet, its branch conventions,
and the four nonzero blocks of the canonical connection. The master integrals are provided in the ancillary file \texttt{basis.m}.

The two-loop alphabet $\{ W_1,\ldots, W_{20}\}$ contains 20 letters. Together with the six letters already used at one loop (see \eqref{one-loop letters 1} and \eqref{one-loop letters 2}), the additional letters are 
\begin{align}
 W_{7}
 &=1-x,
 &
 W_{14}
 &=x(1+y)^2-z^2,
 \notag\\[2pt]
 W_{8}
 &=1+x,
 &
 W_{15}
 &=x+y+xy(x+y)-z^2,
 \notag\\[2pt]
 W_{9}
 &=1-y,
 &
 W_{16}
 &=\frac{(1+x)(Q_1+\Delta)-2z^2}{2z},
 \notag\\[4pt]
 W_{10}
 &=(x-y)^2(1-xy)^2+Q_2^2z^2,
 &
 W_{17}
 &=\frac{(1+x)^2}{2z^2}
   \Bigl[
      (x-y)^2Q_1(Q_1+\Delta)
      +z^2(P_{17}+Q_2\Delta)
   \Bigr],
 \notag\\[4pt]
 W_{11}
 &=\frac{2x(1+x)y-z^2+z\chi}
         {2x(1+x)y-z^2-z\chi},
 &
 W_{18}
 &=\frac{(1+x)
   \bigl[(x+y)(Q_1+\Delta)-2z^2\bigr]}{2z},
 \notag\\[4pt]
 W_{12}
 &=\frac{-2x(1+y)+z^2+z\chi}
         {-2x(1+y)+z^2-z\chi},
 &
 W_{19}
 &=\frac{P_{19}+(1-x)^2z\chi}
         {P_{19}-(1-x)^2z\chi},
 \notag\\[4pt]
 W_{13}
 &=(1+x)^2y-z^2,
 &
 W_{20}
 &=\frac{P_{20}+Q_2z\chi\Delta}
         {P_{20}-Q_2z\chi\Delta}.
 \label{eq:two-loop-additional-letters}
\end{align}
where we recall that our Euclidean region is defined by 
\begin{equation}
 0<x,y<1 \, ,\qquad 0<z<2\sqrt{xy} \, ,\qquad \chi=i\sqrt{4xy-z^2} \, ,\qquad \Delta>0 \, .
\end{equation}
We have also defined
\begin{gather}
 \chi=\sqrt{z^2-4xy},\qquad
 \Delta=\sqrt{(1+x)^2(1+y)^2-4z^2}, \notag\\
 Q_1=(1+x)(1+y),\qquad
 Q_2=(1-x)(1-y), \notag\\
 P_{17}=1-3x^2+4xy-3y^2+x^2y^2, \notag\\
 P_{19}=2x(x-y)(1-xy)+(1-x)^2z^2, \notag\\
 P_{20}=2(1+x)^2y(y-x)(1-xy)
 +\bigl[(1+x^2)(1+y^2)-4(1-x+x^2)y\bigr]z^2.
 \label{eq:two-loop-letter-abbreviations}
\end{gather}
Writing $\omega_\alpha:= \mathrm{d} \log W_\alpha$, the two-loop connection matrix~\eqref{eq:two-loop-block-system} reads as follows. First,
\begin{align}
 \Omega_{10}&=
 \begin{pmatrix}2\omega_{1}\\-\omega_{5}\\-\omega_{2}\end{pmatrix},
 \label{eq:Omega10}\\[3pt]
 \Omega_{21}&=
 \begin{pmatrix}
 -\tfrac12\omega_{3}&-2\omega_{6}&\omega_{4}\\
 4\omega_{1}&0&0\\
 -2\omega_{1}&0&4\omega_{2}\\
 \tfrac12(\omega_{5}-\omega_{11})&-\omega_{13}+\omega_{14}&-\omega_{5}-\omega_{12}\\
 \omega_{1}-\tfrac14\omega_{2}-\tfrac12\omega_{7}-\omega_{8}+\tfrac14\omega_{13}
   &\tfrac12\omega_{11}&-\omega_{2}\\
 -\tfrac54\omega_{5}+\tfrac12\omega_{11}
   &\tfrac32\omega_{1}+\omega_{13}-\omega_{14}&\omega_{5}+\omega_{12}\\
 \omega_{1}-\tfrac12\omega_{2}-\omega_{7}-2\omega_{8}+\tfrac12\omega_{13}
   &-\omega_{5}+\omega_{11}&0
 \end{pmatrix}.
 \label{eq:Omega21}
\end{align}
To keep the \(3\times7\) block legible, we write
\begin{equation}
 \Omega_{32}=
 \begin{pmatrix}
 u_5&u_6&u_7&u_8&u_9&u_{10}&u_{11}\\
 v_5&v_6&v_7&v_8&v_9&v_{10}&v_{11}\\
 w_5&w_6&w_7&w_8&w_9&w_{10}&w_{11}
 \end{pmatrix}.
 \label{eq:Omega32}
\end{equation}
Its entries, ordered by \(j=5,\ldots,11\), are
\begin{equation}
 (u_5,\ldots,u_{11})^{\mathsf T}
 =\begin{pmatrix}
 -2\omega_{5}-\omega_{19}\\
 -\tfrac12\omega_{6}\\
 \omega_{6}+\tfrac12\omega_{20}\\
 8\omega_{8}-\tfrac{10}{3}\omega_{3}+2\omega_{10}-4\omega_{13}
   +8\omega_{16}-4\omega_{17}\\
 8\omega_{6}\\-\tfrac43\omega_{3}\\-4\omega_{6}
 \end{pmatrix},
 \label{eq:u-vector}
\end{equation}
\begin{equation}
 (v_5,\ldots,v_{11})^{\mathsf T}
 =\begin{pmatrix}
 -\tfrac12\omega_{2}-\omega_{7}+\tfrac14\omega_{10}-\tfrac12\omega_{13}
   +\tfrac12\omega_{15}\\
 0\\
 -\tfrac18\omega_{10}+\tfrac14\omega_{15}+\tfrac14\omega_{17}
   -\tfrac12\omega_{18}\\
 -\tfrac12\omega_{20}\\
 4\omega_{8}-\omega_{3}-2\omega_{13}+2\omega_{15}+4\omega_{16}-4\omega_{18}\\
 0\\
 -2\omega_{8}+\omega_{13}-\omega_{15}-2\omega_{16}+2\omega_{18}
 \end{pmatrix},
 \label{eq:v-vector}
\end{equation}
\begin{equation}
 (w_5,\ldots,w_{11})^{\mathsf T}
 =\begin{pmatrix}
 4\omega_{8}-\omega_{3}+\tfrac12\omega_{10}-2\omega_{13}+\omega_{15}
   +4\omega_{16}-\omega_{17}-2\omega_{18}\\
 -\tfrac14\omega_{1}+\omega_{8}+\tfrac12\omega_{9}-\tfrac12\omega_{13}
   +\tfrac14\omega_{14}\\
 \tfrac12\omega_{1}+\omega_{9}-\tfrac14\omega_{10}-\tfrac12\omega_{14}
   +\tfrac12\omega_{15}\\
 \tfrac43\omega_{5}-\tfrac43\omega_{11}-\omega_{19}\\
 -4\omega_{14}+4\omega_{15}\\
 \tfrac43\omega_{5}-\tfrac43\omega_{11}\\
 2\omega_{1}+2\omega_{14}-2\omega_{15}
 \end{pmatrix}.
 \label{eq:w-vector}
\end{equation}
Finally,
\begin{equation}
 \Omega_{43}=
 \begin{pmatrix}-\omega_{5}+\omega_{11}&-4\omega_{1}&\omega_{3}\end{pmatrix}.
 \label{eq:Omega43}
\end{equation}

\subsection{Boundary data}
\label{app:Boundary data}

\paragraph{Configuration $z=0$.}
In the limit $z \rightarrow 0$, the two-loop differential equation system splits into an eleven-dimensional block and a four-dimensional one, with the latter being spanned by the directions $(f_3,f_8,f_{10},f_{12})$. Let
$\boldsymbol g=(g_1,\ldots,g_{11})^{\mathsf T}$ denote the two-loop basis used for the differential equation in~\cite{Flieger:2025ekn}. The identification between the basis $\boldsymbol g$ and the eleven-dimensional block appearing at $z \rightarrow 0$ is given by the linear transformation
\begin{equation}
\boldsymbol g=\bigl(-4f_1,\;4f_4,\;-2f_2,\;-2f_5,\;
 \tfrac12f_6+f_7,
2f_7+8f_9-4f_{11},\;-\tfrac12f_6,\;-f_6+4f_{11},
2f_{14},\;4f_{13},\;-f_{15}\bigr) \, ,
\label{eq:z0-basis-map}
\end{equation}
where the master integrals $f_i$ are evaluated in the limit $z \rightarrow 0$.
This is obtained by writing both bases in the same normalized integral family and by using the IBP relations at $z=0$.
The remaining four integrals can be integrated from the boundary vector above using the $z=0$ DE system at the numerical boundary value $(f_3, f_8, f_{10}, f_{12})|_{(1,1,0)} = (i \pi, 0, 0, 0)$. The first three master integrals are
\begin{equation}
    \begin{aligned}
 f_3(x,y,0)=i\pi \, , \quad
 f_8(x,y,0)=i\pi\log\!\frac{x(1+y)^2}{y(1+x)^2} \, , \quad
 f_{10}(x,y,0)=i\pi\log\!\frac{\sqrt{x}(1+x)^2y}{(1+y)^2} \, .
 \label{eq:extra-boundary-values}
 \end{aligned}
\end{equation}
The last master integral is a pure function of transcendental weight three determined by its differential equation
\begin{equation}
 \dd f_{12}=-4f_8\,\dd\log\!\frac{x-y}{1-xy}
             +4i\pi\log x\,\dd\log y \, ,
 \qquad f_{12}(1,y)=0 \, .
 \label{eq:f12-boundary-equation}
\end{equation}
The explicit expression of $f_{12}(x,y,0)$, as well as the complete fifteen-component boundary vector $\boldsymbol{f}^{(2)}\big|_{(x,y,0)}$, is supplied in the ancillary file \texttt{boundary.m}.

\paragraph{Configuration $z=2\sqrt{xy}$.}
As a further consistency check, we recover the system of \cite{Caron-Huot:2014lda} in the kinematic limit $z \rightarrow 2\sqrt{xy}$, corresponding to $M^2=0$. We approach this limit from the Euclidean
chamber with the branch choice of \eqref{eq:branch-choice}. Pulling back the
connection at fixed $z/\sqrt{xy}$, using the chain rule and
then taking the limit gives:
\begin{equation}
 \left.\dd A^{(2)}\right|_{z=2\sqrt{xy}}
 =\lim_{z\uparrow2\sqrt{xy}}\left[
 \left(\partial_x A^{(2)}+\frac{z}{2x}\partial_z A^{(2)}\right)\dd x
 +\left(\partial_y A^{(2)}+\frac{z}{2y}\partial_z A^{(2)}\right)\dd y
 \right].
 \label{eq:chh-tangent-connection}
\end{equation}
Some entries of the connection matrices vanish in this limit, and we have to accordingly find a map between the minimal set of masters involved in the boundary system, and those of \cite{Caron-Huot:2014lda}. All masters below are evaluated at $(x,y,2\sqrt{xy})$. Four of them vanish, and together with
the product identity $f_7+4f_9-2f_{11}=-f_3^2$, we have
\begin{equation}
 f_3=f_8=f_{10}=f_{12}=0 \, ,
 \qquad f_7+4f_9-2f_{11}=0 \, .
 \label{eq:chh-surface-relations}
\end{equation}
Therefore we are left with a ten-dimensional system. Denoting the basis in~\cite{Caron-Huot:2014lda} by $\boldsymbol g_{\rm CHH}$, the map between the basis is given by
\begin{equation}
 \begin{aligned}
 \boldsymbol g_{\rm CHH}
 =(f_1,\;\tfrac12f_2,\;-f_4,\;\tfrac18f_6,\;
 -\tfrac12f_{11},\;\tfrac12f_5,
 -\tfrac18f_6+f_9-\tfrac12f_{11},\;-\tfrac12f_{13},\;
 -\tfrac14f_{14},\;\tfrac18f_{15}) \, ,
 \end{aligned}
 \label{eq:chh-basis-map}
\end{equation}
where we suppressed dependence on their two variables $u_{\rm CHH}$ and $\ v_{\rm CHH}$, which are related to our cross ratios as $u_{\rm CHH}=4x/(1-x)^2, \ v_{\rm CHH}=4y/(1-y)^2$.
As a consequence, our $x$ and $y$ connection matrices reproduce Eq.~(4.11) of \cite{Caron-Huot:2014lda}. 

\subsection{GPL representation}\label{app:gpltwoloop}

In this subsection we rationalize the two square roots and write the transport from the $z=0$ boundary to a generic point in the Euclidean region as GPLs of weight at most four. To this end, we introduce two auxiliary rationalization variables $u$ and $v$,
\begin{equation}
 u=\frac{z+\chi}{2x} \, , \qquad
 v=\frac{2z}{\Delta+(1+x)(1+y)} \, .
 \label{eq:rational-variables}
\end{equation}
Here $\operatorname{Im}\chi>0$ and $\Delta>0$ are the same as \eqref{eq:two-loop-letter-abbreviations}. Let $S_1,S_2,S_3$ be three points 
\begin{equation}
 S_1=(x,y,0) \, ,\qquad S_2=(x,-xu^2,0) \, ,\qquad S_3=(x,y,z) \, .
\end{equation}
Keeping
$x$ and $u$ fixed, let us define a path
$\widetilde\gamma(\tau)=(x,y(\tau),z(\tau))$ by
\begin{equation}
\begin{aligned}
 D(\tau)&=1-u\tau(1+x)+\tau^2 \, ,\\
 y(\tau)&=\frac{u\bigl[\tau(1+x)-u x(1+\tau^2)\bigr]}{D(\tau)} \, ,&
 z(\tau)&=\frac{\tau(1+x)(1-u^2x)}{D(\tau)} \, .
\end{aligned}
 \label{eq:rational-path}
\end{equation}
This path goes from $S_2$ to $S_3$ as $\tau$ runs from $0$ to $v$, and rationalizes the two roots:
\begin{equation}
 \chi(\tau)=
 \frac{2 u x(1+\tau^2)-\tau(1+x)(1+u^2x)}{D(\tau)} \, ,
 \qquad
 \Delta(\tau)=
 \frac{(1-\tau^2)(1+x)(1-u^2x)}{D(\tau)} \, .
 \label{eq:rationalized-roots}
\end{equation}
The direct $z$-integration transports $S_1\rightarrow S_3$ along
$\gamma(t)=(x,y,tz)$. The rationalized integration path instead follows
$S_1\rightarrow S_2\rightarrow S_3$, where the first transport is along
\begin{equation}
 \beta(s)=\bigl(x,(1-s)y-sxu^2,0\bigr) \, ,\qquad 0\leq s\leq1 \, ,
 \label{eq:boundary-continuation-path}
\end{equation}
and the second is along $\tilde\gamma(\tau)$.
Flatness of $\dd A^{(2)}$ makes the two routes equivalent, since one can be deformed into the other without passing through a singularity of $A^{(2)}$.

The pulled-back connection has 26 simple poles $a_j(x,u)$,
\begin{equation}
 \widetilde\gamma^{*}(\dd A^{(2)})
 =\sum_{j=1}^{26}\widehat R_j\frac{\dd\tau}{\tau-a_j} . \label{eq:gpl-definition}
\end{equation}
where each $\widehat R_j$ is a constant rational linear combination of the
alphabet matrices $R_\alpha$ appearing in $A^{(2)}$ in the two-loop equivalent of the canonical form \eqref{eq:canform}. As a consequence of this pole structure, iterated integrals contributing to the double-box can be expressed in terms of Goncharov polylogarithms, defined as
\begin{equation}
    G(b_1,\ldots,b_n;v)
 =\int_0^v\frac{\dd\tau}{\tau-b_1}
   G(b_2,\ldots,b_n;\tau) \, , \quad
 G(\,;v)=1 \,.
\end{equation}
Here each index $b_i$ is one of the poles $a_j(x,u)$ of the pulled-back
connection, while $v$ is the endpoint of the rationalized path. The number
$n$ is the transcendental weight. We also use the notation $G(0,...,0;v) = \log^n(v)/n!$.
The complete result is supplied in the ancillary file \texttt{doublebox.m}. 

From~\eqref{eq:dbtoamp} and~\eqref{eq:f15norm}, we obtain 
\begin{equation}
 \mathcal{B}_s^{(2)}(s,t;m,M)
 =\frac{(1+x^2)^2(1+y^2)}{(1-x)^2(1+x)(1-y)\Delta}\,f_{15}(x,y,z) \, .
 \label{eq:f15-to-B2}
\end{equation}
Thus, the proposed two-loop amplitude is expressible in terms of ordinary GPLs with algebraic letters and prefactors. The function \texttt{B2[x,y,z]} in \texttt{doublebox.m} returns the symbolic GPL expression for $\mathcal{B}_s^{(2)}(s,t;m,M)$; \texttt{B2[y,x,z]} returns the crossed channel expression.

\section{Walking asymptotics}\label{app:walking}

This appendix derives the one- and two-loop walking coefficients~\eqref{eq:G1} and~\eqref{eq:G2} presented in Section \ref{sec:walking}.

\subsection{One-loop extraction}

Let us begin the computation of the one-loop walking anomalous dimension $\Gamma_{\rm walk}^{(1)}$ by studying the IR behavior of the one-loop letters introduced in \eqref{one-loop letters 1} and \eqref{one-loop letters 2}.
Let us recall that for \(\rho>0\) we choose \(0<\kappa_s\leq\kappa_t\), while for \(\rho<0\) we take \(0<\kappa_t\leq\kappa_s\). Then, 
the logarithmic slopes of the six one-loop letters, as defined in \eqref{IR letters}, are
\begin{equation}
 \begin{aligned}
 (-\nu_1,\ldots,-\nu_6)=\bigl(&
 -\max(\kappa_s+\rho,0),-\max(\kappa_t+\rho,0),
2\max(\kappa_t+\rho-\min(\kappa_s+\rho,0),0)\\[-2pt]
&-2\max(\rho-\kappa_s,0),
 2\max(\kappa_s+\rho-\min(\kappa_t+\rho,0),0)-2\max(\rho-\kappa_t,0),\\[-2pt]
 &-2 \max(\rho,0),
  \min(2 \max(\rho,0),\kappa_s+\max(\rho,0),\kappa_t+\max(\rho,0),\kappa_s+\kappa_t)
 \bigr) \, .
 \end{aligned}
 \label{eq:app-one-loop-slopes}
\end{equation}
By reading the coefficient of each $\mathrm d\log W_\alpha$ from \eqref{eq:one-loop-canonical-system} we can obtain the $R_{\rm box}$ matrix that dictates the one-loop differential equation \eqref{eq:one-loop-extraction} in the IR, which gives
\begin{equation}
 R_{\rm box}=
 \begin{pmatrix}
 0&0&0&0&0\\
 -2\nu_1&0&0&0&0\\
 \nu_5&0&0&0&0\\
 \nu_2&0&0&0&0\\
 0&\tfrac12\nu_3&2\nu_6&-\nu_4&0
 \end{pmatrix}.
 \label{eq:app-one-loop-residue}
\end{equation}
As discussed in Section \ref{sec:walking}, we compute the leading IR scaling of the one-loop master integrals $\boldsymbol f^{(1)}$ by solving the differential equation \eqref{eq:residue} as in \eqref{eq:nilpotenttransport}, for which we need to determine the asymptotic boundary vector $\boldsymbol f^{(1)}_0$ defined in \eqref{f0 walking}. To determine these constants, we continue the known boundary solution from $z=0$ to $z=z(a)$ at fixed $x(a)$ and $y(a)$, deforming the real integration segment into the upper half-plane to avoid singularities. Square roots and logarithms are transported continuously from their boundary branches, following the prescription in Section~\ref{sec:walking}. Taking the limit \eqref{f0 walking} gives
\begin{align}
\boldsymbol f^{(1)}_0=\begin{cases}
\left(1,0,0,0,0 \right), & \rho<-\kappa_t \,, \\
\left(1,0,0,0,-\pi^2 \right), & -\kappa_t<\rho<0 \,, \\
\left(1,0,0,0,-\pi^2 \right), & 0<\rho<\kappa_s \,, \\
\left(1,0,0,0,-\tfrac{\pi^2}{3} \right), & \kappa_s<\rho<\kappa_t \,, \\
\left(1,0,0,0,\tfrac{\pi^2}{3}  \right), & \rho>\kappa_t \,. 
\end{cases}
\label{f0 one loop}
\end{align}
Plugging \eqref{f0 one loop} into \eqref{eq:nilpotenttransport} we obtain
\begin{equation}
 [a^2](-\mathcal M^{(1)})
 =\frac12\bigl(R_{\rm box}^2\boldsymbol e_1\bigr)_5 \, ,
 \ \ 
 \bigl(R_{\rm box}^2\bigr)_{51}
 =R_{52}R_{21}+R_{53}R_{31}+R_{54}R_{41}
 =-\nu_1\nu_3+2\nu_5\nu_6-\nu_2\nu_4 \, .
 \label{eq:app-one-loop-chain}
\end{equation}
with $\boldsymbol e_1=(1,0,0,0,0)^{\mathsf T}$.
Let us stress that, since $R_{\rm box}^3=0$,
from \eqref{eq:nilpotenttransport} the leading IR behavior of ${\cal M}^{(1)}$ (and therefore of $\log{\cal M}$ at one loop) contains no $a^3$ or $a^4$ terms.  
As we will see in the following section, those scalings are allowed in the two-loop amplitude ${\cal M}^{(2)}$, and their absence in $\log{\cal M}$ at two loops requires a precise cancellation between ${\cal M}^{(2)}$ and $({\cal M}^{(1)})^2$, as follows from \eqref{eq:logexpansion}.
Substitution of \eqref{eq:app-one-loop-slopes} in \eqref{eq:app-one-loop-chain} gives 
\begin{equation}
 [a^2](-\mathcal M^{(1)})=
 \begin{cases}
 0 \, ,&\rho\leq-\kappa_t \, ,\\
 2(\kappa_s+\rho)(\kappa_t+\rho) \, ,
   &-\kappa_t<\rho\leq0 \, ,\\
 2\kappa_s\kappa_t+2\rho(\kappa_s+\kappa_t)-2\rho^2 \, ,
   &0<\rho\leq\kappa_s \, ,\\
 (\kappa_s+\rho)(\kappa_s+2\kappa_t-\rho) \, ,
   &\kappa_s<\rho\leq\kappa_t \, ,\\
 (\kappa_s+\kappa_t)^2,&\rho>\kappa_t \, .
 \end{cases}
 \label{eq:app-one-loop-coefficient}
\end{equation}
Dividing by $\kappa_s\kappa_t$ and using
$\eta_s=\rho/\kappa_s$ and $\eta_t=\rho/\kappa_t$ we obtain \eqref{eq:G1}.

\subsection{Two-loop extraction}
\label{app:two-loop walking}

We now turn to the computation of the two-loop walking anomalous dimension $\Gamma_{\rm walk}^{(2)}$. The computation follows essentially the same steps as in the one-loop analysis presented in the previous section. The main difference is that in order to compute the IR behavior of the two-loop amplitude ${\cal M}^{(2)}$ we need to study the IR scaling of both the s-channel and t-channel double boxes. Consequently, we need to study the IR limit of the two-loop differential equation discussed in Section \ref{sec:2loopresult} but also the asymptotics of the corresponding t-channel system, obtained by a $x \leftrightarrow y$ permutation.

Let us begin by presenting the s-channel letter slopes. The corresponding t-channel slopes can accordingly be computed by a $\kappa_s \leftrightarrow \kappa_t$ permutation. The first six letters behave as in \eqref{eq:app-one-loop-slopes}, while the remaining slopes $(-\nu_7,\ldots,-\nu_{20})$ read
\begin{equation}
\begin{aligned}
\Bigl[
&\frac{\min(\kappa_s+\rho,0)}{2},\;
0,\;
\frac{\min(\kappa_t+\rho,0)}{2},
\\[2pt]
&-2\min\!\left(
  \frac{\kappa_s+\kappa_t}{2},
  \max(\kappa_s+\rho,0),
  \max(\kappa_t+\rho,0)
  \right)
  +2\min\!\bigl(\max(\kappa_s+\rho,\kappa_t+\rho),0\bigr),
\\[2pt]
&-2\max(\rho,0)
 -\min\!\bigl(
   2\max(\rho,0),
   \kappa_s+\max(\rho,0)
   \bigr),
\qquad
\max(0,\rho-\kappa_t),
\\[2pt]
&\min\!\bigl(\max(\kappa_s+\rho,\kappa_t+\rho),0\bigr)
 -\max\!\bigl(
   \kappa_t+\rho-\min(\kappa_s+\rho,0),0
   \bigr)
 +\max(\rho-\kappa_s,0),
\\[2pt]
&\min\!\bigl(\max(\kappa_s+\rho,\kappa_t+\rho),0\bigr)
 -\max\!\bigl(
   \kappa_s+\rho-\min(\kappa_t+\rho,0),0
   \bigr)
 +\max(\rho-\kappa_t,0),
\\[2pt]
&\min\!\bigl(\max(\kappa_s+\rho,\kappa_t+\rho),0\bigr)
 -\max\!\bigl(\min(\kappa_s+\rho,\kappa_t+\rho),0\bigr)
 +\max\!\bigl(\rho-\max(\kappa_s,\kappa_t),0\bigr),
\\[2pt]
&\frac{
  \max(\kappa_s+\rho,0)
  +\max(\kappa_t+\rho,0)
  +\min\!\bigl(\max(\kappa_s+\rho,\kappa_t+\rho),0\bigr)
  -2\max(\rho,0)
  }{2},
\\[2pt]
&\max\!\left(
  0,
  \left|
  \max(\kappa_s+\rho,0)-\max(\kappa_t+\rho,0)
  \right|
  -2\max(\rho,0)
  \right)
  +\min\!\bigl(\max(\kappa_s+\rho,\kappa_t+\rho),0\bigr),
\\[2pt]
&\frac{
  \left|
  \max(\kappa_s+\rho,0)-\max(\kappa_t+\rho,0)
  \right|
  +\min\!\bigl(\max(\kappa_s+\rho,\kappa_t+\rho),0\bigr)
  }{2}
\\[-1pt]
&\hspace{2em}
  +\min(\kappa_s,\kappa_t)
  -\min\!\bigl(
    \kappa_s+\max(\rho,0),
    \kappa_t+\max(\rho,0),
    \kappa_s+\kappa_t
    \bigr),
\\[2pt]
&\min\!\bigl(
  4\max(\rho,0),
  \max(0,\kappa_s-\kappa_t+2\max(\rho,0))
  \bigr),
\\[2pt]
&2\min\!\bigl(
  2\max(\rho,0),
  \kappa_s+\max(\rho,0)
  \bigr)
  -\min\!\bigl(
    4\max(\rho,0),
    \max(0,\kappa_s-\kappa_t+2\max(\rho,0))
    \bigr)
\Bigr].
\end{aligned}
\label{eq:app-two-loop-slopes}
\end{equation}
The residue matrices $R_{\alpha}$ in~\eqref{eq:residue} are read entry by entry from the block connection in~\eqref{eq:Omega10}--\eqref{eq:Omega43} of
Appendix~\ref{app:twoloopexpr}:
\begin{equation}
 R_\alpha=[\omega_\alpha]
 \begin{pmatrix}
 0&0&0&0&0\\
 \Omega_{10}&0&0&0&0\\
 0&\Omega_{21}&0&0&0\\
 0&0&\Omega_{32}&0&0\\
 0&0&0&\Omega_{43}&0
 \end{pmatrix},
 \qquad \omega_\alpha=\mathrm d\log W_\alpha,
 \label{eq:app-residue-extraction}
\end{equation}
where $[\omega_\alpha]$ means taking the coefficient of
$\omega_\alpha$ in every entry.  Combining these matrices with the slopes
above gives the walking ray residues
\begin{equation}
 R_s=-\sum_{\alpha=1}^{20}\nu_\alpha(\kappa_s,\kappa_t,\rho)R_\alpha \, ,
 \qquad
 R_t=-\sum_{\alpha=1}^{20}\nu_\alpha(\kappa_t,\kappa_s,\rho)R_\alpha \, .
 \label{eq:app-two-loop-residues}
\end{equation}
where $R_s$ and $R_t$ are respectively the matrices of the s-channel and t-channel differential equations in their IR limit.
Thus,~\eqref{eq:app-two-loop-slopes}--\eqref{eq:app-two-loop-residues}
and the blocks in Appendix~\ref{app:twoloopexpr} determine $R_s$ and $R_t$.

As at one loop, the extraction of \(\Gamma_{\rm walk}^{(2)}\) requires the asymptotic boundary vector \(\boldsymbol f_0^{(2)}\). Only its first eleven components contribute, and the relevant weight-two entries are
\begin{equation}
 \frac{\bigl((\boldsymbol f^{(2)}_{0,h})_5,\ldots,
 (\boldsymbol f^{(2)}_{0,h})_{11}\bigr)}{\pi^2}=
 \begin{cases}
 (0,0,0,0,0,0,0) \, ,& \rho<-\kappa_s \,,\\
 (0,0,0,0,-\tfrac16,0,-\tfrac13) \, ,&-\kappa_s<\rho<-\kappa_t \,, \ h=s\,,\\
 (0,0,0,0,0,0,0) \, ,&-\kappa_s<\rho<-\kappa_t \,, \ h=t\,,\\
 (-1,0,0,0,-\tfrac16,0,-\tfrac13) \, ,&-\kappa_t<\rho<0 \,,\\
 (-1,0,0,0,-\tfrac16,0,-\tfrac13) \, ,&0<\rho<\kappa_s \,,\\
 (-\tfrac13,0,0,-\tfrac13,0,\tfrac13,0) \, ,
     &\kappa_s<\rho<\kappa_t,\ h=s \,,\\
 (-\tfrac13,0,0,\tfrac13,-\tfrac16,-\tfrac13,-\tfrac13) \, ,
     &\kappa_s<\rho<\kappa_t,\ h=t \,,\\
 (\tfrac13,0,0,0,0,0,0) \, ,&\rho>\kappa_t \, .
 \end{cases}
 \label{eq:app-weight-two-matching}
\end{equation}

Combining these boundary data with the residue matrices in~\eqref{eq:app-two-loop-residues}, the quartic and cubic terms in the logarithm of the amplitude cancel,
\begin{equation}
 [a^4]\left[\mathcal M^{(2)}
       -\frac12(\mathcal M^{(1)})^2\right]
 =
 [a^3]\left[\mathcal M^{(2)}
       -\frac12(\mathcal M^{(1)})^2\right]
 =0\,,
 \label{eq:app-two-loop-exponentiation}
\end{equation}
demonstrating infrared exponentiation through two loops. In deriving this result we used that the normalizations in~\eqref{eq:canformint} and~\eqref{eq:canformint2} approach unity in the infrared for $\kappa_s+\rho>0$ and $\kappa_t+\rho>0$. The only exception is the region \(\rho\leq-\kappa_t\), where the normalizations may diverge. Nevertheless, for $\rho\leq-\kappa_t$ the normalized integrals give at most a linear growth in $a$ for $\mathcal M^{(2)}-\tfrac12(\mathcal M^{(1)})^2$, so also its quadratic coefficient vanishes.

The remaining quadratic term is
\begin{equation}
 [a^2]\left[\mathcal M^{(2)}
       -\frac12(\mathcal M^{(1)})^2\right]
 =\frac{2\pi^2}{3}
 \begin{cases}
 0 \, ,&\rho\leq-\kappa_t \,,\\
 (\kappa_s+\rho)(\kappa_t+\rho) \, ,
    &-\kappa_t<\rho\leq0 \,,\\
 (\kappa_s+\rho)(\kappa_t+\rho) \, ,
    &0<\rho\leq\kappa_s \,,\\
 (\kappa_s+\rho)(\kappa_s+\kappa_t) \, ,
    &\kappa_s<\rho\leq\kappa_t \,,\\
 (\kappa_s+\kappa_t)^2 \, ,&\rho>\kappa_t \, .
 \end{cases}
 \label{eq:app-connected-coefficient}
\end{equation}
Dividing by \(-\kappa_s\kappa_t\) and substituting
\(\eta_s=\rho/\kappa_s\) and \(\eta_t=\rho/\kappa_t\) reproduces~\eqref{eq:G2}. To obtain \eqref{eq:app-connected-coefficient} we assumed that the relative normalization between \(\mathcal B^{(1)}\) and \(\mathcal B_s^{(2)}+\mathcal B_t^{(2)}\) and the amplitudes \(\mathcal M^{(1)}\) and \(\mathcal M^{(2)}\) is the same as in the \(z=0\) Coulomb-branch configuration~\cite{Arkani-Hamed:2023epq,Flieger:2025ekn}. The exponentiation shown in \eqref{eq:app-two-loop-exponentiation} supports this assumption. Finally, let us note that the vectors in \eqref{eq:app-weight-two-matching} apply within the open kinematic chambers. The expressions of the final walking coefficients on the corresponding boundaries are understood as continuous extensions from the adjacent chambers.

\newpage
\bibliographystyle{JHEP}
\bibliography{ref}

\end{document}